\documentclass{article}

\usepackage{arxiv}
\usepackage{graphicx}
\usepackage{multirow}
\usepackage{amsmath,amssymb,amsfonts}
\usepackage{amsthm}
\usepackage{mathrsfs}
\usepackage[title]{appendix}
\usepackage{xcolor}
\usepackage{textcomp}
\usepackage{manyfoot}
\usepackage{booktabs}
\usepackage{algorithm}
\usepackage{algorithmicx}
\usepackage{algpseudocode}
\usepackage{listings}
\usepackage[font=bf]{caption}
\usepackage[section]{placeins}
\usepackage{hyperref}       
\usepackage[square,comma,numbers,sort&compress]{natbib}

\graphicspath{{figs/}}

\newcommand{\yr}{2024 YR$_4$}
\newcommand{\yrs}{2024 YR$_4\;$}

\hypersetup{
pdftitle={Operational Mass Measurement for Flyby Reconnaissance Missions of Potentially Hazardous Asteroids},
pdfsubject={astro-ph.IM, astro-ph.EP, physics.space-ph},
pdfauthor={Justin A. Atchison, Gael Cascioli, Anivid Pedros-Faura, Erwan Mazarico, Rylie A. Bull, Jay McMahon, Evan J. Smith, Daniel R. Cremons},
pdfkeywords={asteroid mass, asteroid gravity science, asteroid flyby science, rapid reconnaissance},
}

\title{Operational Mass Measurement for Flyby Reconnaissance Missions of Potentially Hazardous Asteroids}

\renewcommand{\shorttitle}{Operational Mass Measurement for Flyby Recon. Missions of PHAs}

\author{Justin A. Atchison\\
	Johns Hopkins Univ. Applied Physics Laboratory\\
    11100 Johns Hopkins Road\\
    Laurel, MD 20723 USA\\
	\textit{justin.atchison@jhuapl.edu} \\
	\And
	Gael Cascioli\\
    Univ. of Maryland Baltimore County\\
    1000 Hilltop Circle\\
    Baltimore, MD 21250 USA\\
    \& Goddard Space Flight Center\\
    8800 Greenbelt Road \\
    Greenbelt MD, 20771 USA\\
    \And
    Anivid Pedros-Faura\\
    University of Colorado Boulder\\
    3775 Discovery Drive\\
    Boulder, CO 80309 USA\\
    \And
    Erwan Mazarico\\
    Goddard Space Flight Center\\
    8800 Greenbelt Road \\
    Greenbelt MD, 20771 USA\\
    \And
    Rylie A. Bull\\
	Johns Hopkins Univ. Applied Physics Laboratory\\
    11100 Johns Hopkins Road\\
    Laurel, MD 20723 USA\\    
    \And
    Jay McMahon\\
    University of Colorado Boulder\\
    3775 Discovery Drive\\
    Boulder, CO 80309 USA\\    
    \And 
    Evan J. Smith\\
	Johns Hopkins Univ. Applied Physics Laboratory\\
    11100 Johns Hopkins Road\\
    Laurel, MD 20723 USA\\    
    \And 
    Daniel R. Cremons\\
    Goddard Space Flight Center\\
    8800 Greenbelt Road \\
    Greenbelt MD, 20771 USA\\    
}

\begin{document}

\maketitle

\begin{abstract}
This study evaluates a technique for determining the mass of a potentially hazardous asteroid from a high-speed flyby in the context of a rapid reconnaissance planetary defense scenario. We consider a host spacecraft that dispenses a small CubeSat, which acts as a test-mass. Both spacecraft perform approach maneuvers to target their flyby locations, with the host targeting a close proximity flyby and the CubeSat targeting a distant flyby. By incorporating short-range intersatellite measurements between the host and the CubeSat, the mass measurement sensitivity is substantially improved. We evaluate a set of proposed host and CubeSat hardware options against the 2023 and 2025 Planetary Defense Conference hypothetical threats, as well as a hypothetical flyby of \yr. These scenarios differ predominantly in their flyby speeds, which span from 1.7 to 22 km/s. Based on these scenarios, we demonstrate that a typical radio-frequency intersatellite measurement is ineffective for asteroids with diameters relevant to planetary defense (\textit{i.e.}, 50 - 500 m). However, we find that augmenting the system with a laser-based intersatellite ranging system or a high-precision Doppler system can enable mass measurements of asteroids as small as 100 m across all cases, and as small as 50 m for the slower ($\leq$ 8 km/s) cases. The results are very sensitive to the timing of the final maneuver, which is used to target the low-altitude flyby point. This presents an operational challenge for the smallest objects, where optical detection times are comparatively late and the optical navigation targeting knowledge converges too slowly.
\end{abstract}

\keywords{asteroid mass \and asteroid gravity science \and asteroid flyby science \and rapid reconnaissance}

\section{Introduction}\label{sec:intro}
In a planetary defense scenario, a potentially hazardous asteroid's (PHA) mass is typically poorly known, despite being a characterization priority (\textit{e.g.}, \cite{Council2018, miller2015, hirabayashi2024}). The asteroid's mass determines the potential impact effects at Earth \cite{Wheeler2024} and the momentum required by a mitigation mission to deflect it \cite{Frazier2024}. For asteroids of the size scale relevant to planetary defense (\textit{i.e.}, $\sim$50-500 m \cite{Smpag2017, Nitep2021}), the mass can only be directly measured \textit{in-situ} with a dedicated reconnaissance spacecraft that performs either a high-speed flyby or an extended rendezvous. Flyby spacecraft missions can reach PHAs faster \cite{Chabot2024}, but cannot currently measure their mass using radio-frequency (RF) tracking from Earth, the current state of practice. The small masses of PHAs require unreasonably or impossibly close flyby distances to produce an acceleration observable in the RF tracking data. This capability gap limits the value of rapid reconnaissance flyby missions \cite{barbee2021} and, if flyby missions are the only viable option, it means that decision makers are missing critical information in a planetary defense scenario \cite{Council2018}. 

\begin{figure}[!thbp]
    \begin{center}
        \includegraphics[width=0.7\columnwidth]{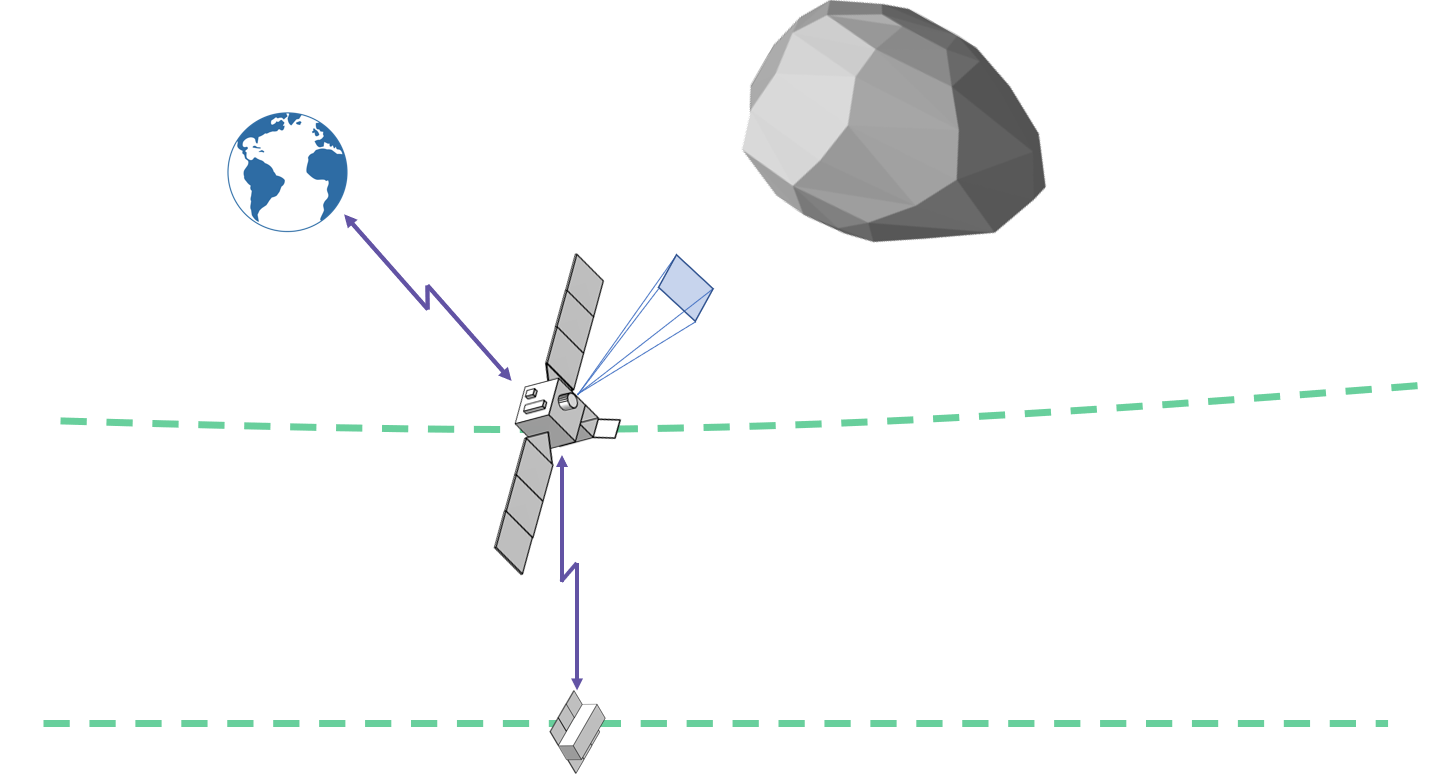}
        \caption{Illustration of flyby intersatellite mass measurement concept, where host spacecraft passes close to the asteroid and cubesat test-mass passes is distant}
        \label{fig:ov1_sketch}
    \end{center}
\end{figure}

Recent research has identified that relative tracking between multiple spacecraft can significantly increase reconnaissance flyby mass measurement sensitivity (\textit{e.g.}, \cite{Christensen2021, Bull2021, Walker2021, park2025, bull2025}). In these approaches, one or more of the spacecraft is targeted to fly very close to the asteroid to maximize the asteroid's acceleration acting on it at close approach. In this paper's scenario, illustrated in Figure~\ref{fig:ov1_sketch}, the flyby spacecraft deploys a trackable test-mass prior to the flyby. The test-mass is intended to be relatively simple and pass roughly 10 kilometers from the asteroid, while the more capable host spacecraft targets a very low altitude flyby (nominally 1-3 body radii). The test-mass's trajectory is unperturbed and serves as a ballistic reference to compare the host's trajectory against. The relatively short distance between the host spacecraft and the test-mass facilitates high accuracy intersatellite measurements, which when coupled with the low flyby provide a very sensitive mass measurement. 

This study evaluates a scenario with a single test-mass, which consists of a maneuverable, three-axis controlled CubeSat. The host is equipped with a camera for optical navigation. Both the host and test-mass spacecraft are equipped with RF radios, such that the host can generate an intersatellite range and Doppler measurement. We also consider scenarios where the the host satellite has an additional laser ranging instrument (LRI) and where the pair have a high precision Doppler (HPD) instrument. 

This concept is modeled in the context of three reference scenarios given in Table~\ref{tab:scenario_list}. The first two are from the 2023 and 2025 Planetary Defense Conference (PDC) hypothetical scenarios. The final is a hypothetical flyby of real asteroid \yr. These scenarios span relatively slow ($<$2 km/s) to fast ($>$20 km/s) flyby speeds. Although we vary asteroid diameter as a parameter, Table~\ref{tab:scenario_list} includes reference sizes for each scenario. 

This study expands on prior work \cite{Atchison2017a, Bull2024} by modeling the full encounter concept-of-operations with representative spacecraft models and realistic measurement and maneuver schedules. We model ground tracking throughout cruise and flyby operations, optical navigation (when the asteroid is detectable), pointing constraints, and multiple approach targeting maneuvers. 

\begin{table}[!ht]
    \centering
    \caption{Simulated Planetary Defense Scenarios}
    \label{tab:scenario_list}
    \begin{tabular}{lrrc}
        \toprule
        Asteroid   & Ref. Diameter & Flyby Speed & Epoch\\
        \midrule
        2023 PDC   & 800 m      & 1.7 km/s  & 01 Dec 2025 \\
        2024 PDC25 & 150 m      & 8.1 km/s  & 12 Apr 2028 \\
        \yr        & 60$\pm$7 m & 22.0 km/s & 11 Nov 2028 \\
        \bottomrule
    \end{tabular}
\end{table}

This paper is organized as follows. In Section~\ref{sec:background} we briefly give rationale for a proposed mass measurement accuracy metric. We also summarize the orders-of-magnitude for the perturbations that we are attempting to measure. In Section~\ref{sec:conops}, we describe the overall concept-of-operations and the associated simulation parameters and measurement model parameters. Sections~\ref{sec:case1}-\ref{sec:case3} each describe the scenario setup and simulation results for the three hypothetical scenarios. Section~\ref{sec:summary} summarizes the full set of results across the scenarios. Finally, Section~\ref{sec:conclusions} concludes and suggests options to improve the results.    

\section{Background}\label{sec:background}
To help motivate this study, we briefly describe the assumed knowledge of an asteroid's mass at the time of its discovery and propose an accuracy metric that would represent a meaningful improvement for planetary defense purposes. We then demonstrate how challenging this metric is by defining the parameters that are being measured.

\subsection{Mass Measurement Metric}
There are currently no defined requirements for mass measurement accuracy in a planetary defense context. The utility of mass data depends on both the application and on the size of the object. For example, mass uncertainty has different implications among the different mitigation techniques and has different levels of significance in Earth threat models when considering a 50 m threat versus a 500 m threat. In an attempt to broadly generalize the needed accuracy for the purposes of this study, we propose to use 25\% (1$\sigma$) as a minimum success criteria. 

\begin{figure}[!thbp]
    \begin{center}
        \includegraphics[width=0.5\columnwidth]{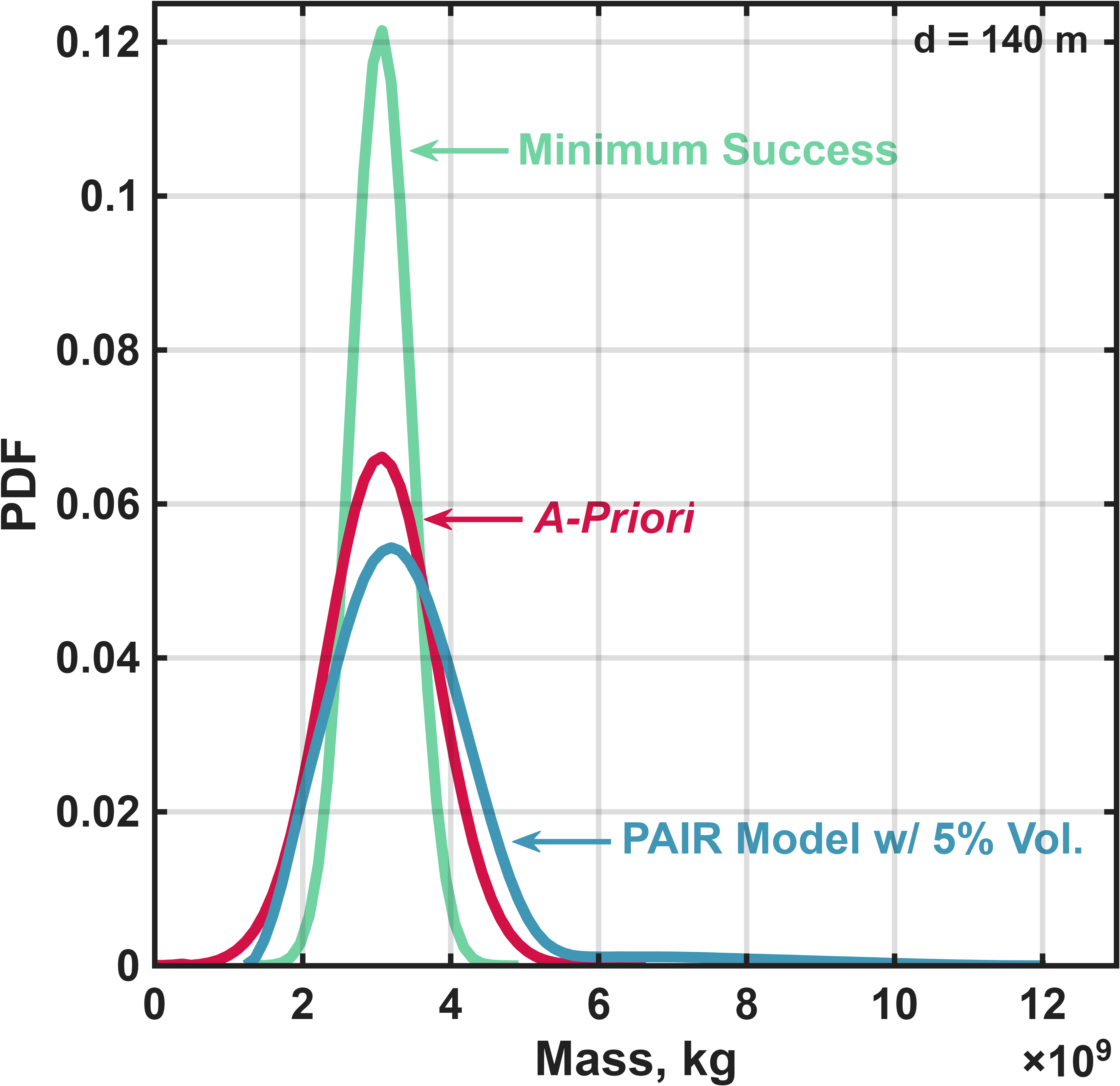}
        \caption{Probability distribution functions for the mass of a 140 m diameter asteroid, including the PAIR model \cite{Mathias2017} and the metrics used in this study}
        \label{fig:mass_pdf}
    \end{center}
\end{figure}

In the current state of practice, when no mass measurement is available, the Probabilistic Asteroid Impact Risk (PAIR) model \cite{Mathias2017} is used by the International Asteroid Warning Network \cite{Kofler2019} to assess the range of threats that a new PHA represents. The model derives a probability distribution of PHA bulk density based on measured meteorites, their relative frequencies, and macroporosity measurements. Coupled with the volume, we can generate a range of expected PHA masses. We expect to constrain the total volume to roughly 5\% $1\sigma$ \cite{barbee2025,Chabot2024} from a high-speed flyby with favorable lighting conditions. Figure~\ref{fig:mass_pdf} shows the resulting range of masses the PAIR model predicts for a 140 m object. The model includes a long tail that extends out to higher masses with a bulk density of up to 7.5 g/cm$^3$ ($\sim$12$\times10^9$ kg with a 3$\sigma$ high volume). This accounts for denser metallic PHAs. The figure also includes our initial modeled \textit{a-priori} uncertainty of 50\% $1\sigma$ (shown in red) and our proposed success metric of 25\% $1\sigma$ shown in green. We propose that the 25\% minimum success metric represents a substantial improvement operationally because it eliminates (or confirms) the relatively low probability risk of a metallic object, and simplifies the range of likely masses by a factor of $\sim$6. We note that while spectral instruments can also confirm that the \textit{surface} of the object is not metallic, a shape-based mass estimate still requires an assumption that the interior has the same composition. Here, we are attempting a direct, unambiguous measurement. 

\subsection{Direct Mass Measurement}
In a flyby encounter, the asteroid's mass can be observed as a very small change in the spacecraft's heliocentric velocity ($\Delta$v), as in a gravitational assist:
\begin{equation}
    \Delta \textrm{v} = \frac{2 \, \textrm{v}_{ca} \, G M}{r_{ca} \, \textrm{v}_{ca}^2 + G\, M} \approx \frac{2 \,GM}{r_{ca} \, \textrm{v}_{ca}}
\end{equation}
\\
where $G$ is the universal gravitational constant, $M$ is the the asteroid's mass, $r_{ca}$ is the distance to the asteroid's center-of-mass at closest approach, and $\textrm{v}_{ca}$ is the velocity relative to the asteroid at closest approach. The simplification results from the observation that $r_{ca} \, \textrm{v}_{ca}^2 >> G\,M$.

The asteroid's sphere-of-influence varies by asteroid size and solar distance, but is on the order of 5-15 body radii (100's of meters to about 5 kilometers) for the scenarios in this study. With flyby speeds on the order of kilometers per second, a spacecraft is only within the asteroid's sphere-of-influence for a few seconds at most. That is, the imparted $\Delta$v is essentially impulsive.

\begin{figure}[!thbp]
    \begin{center}
        \includegraphics[width=0.5\columnwidth]{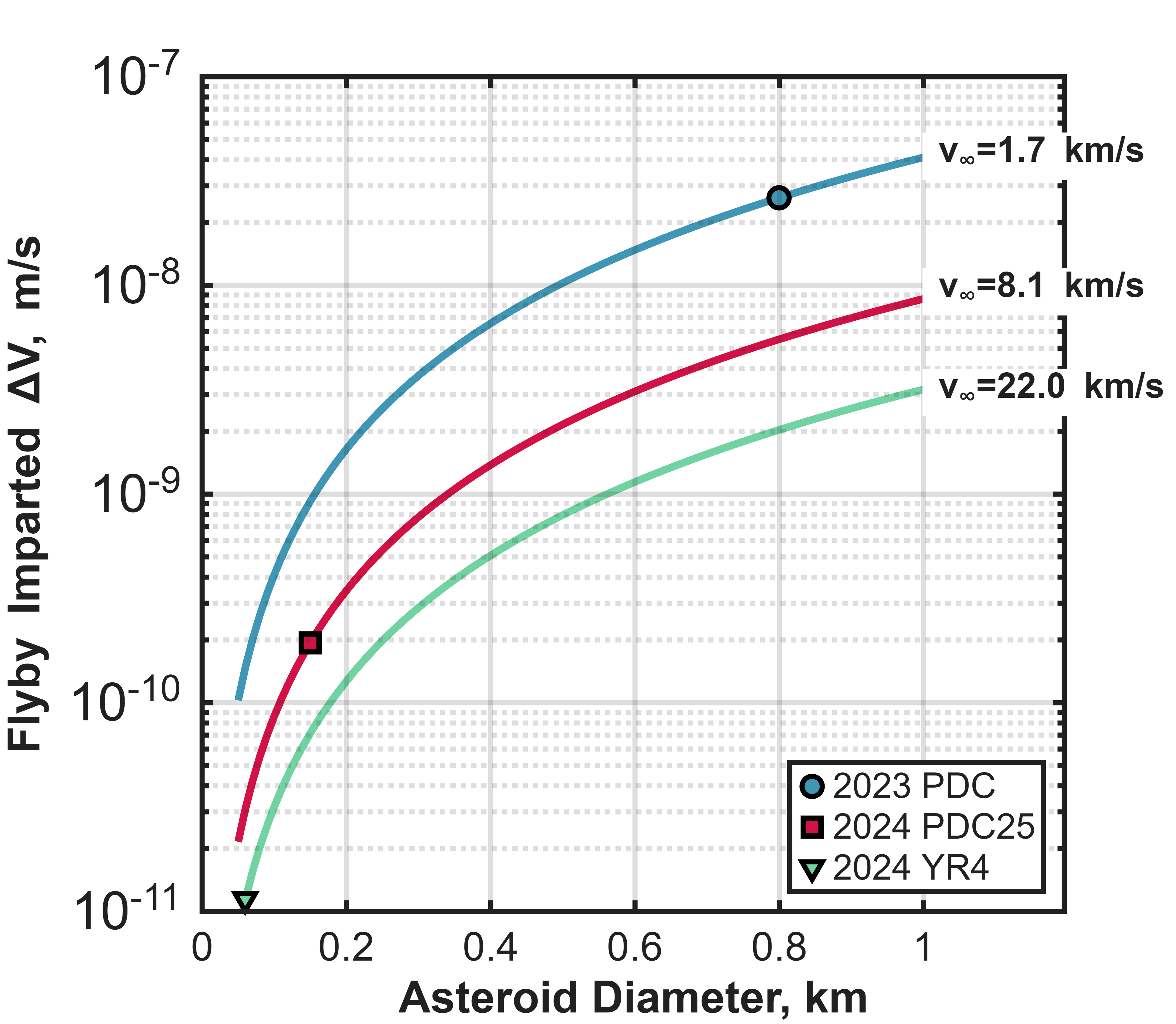}
        \caption{The velocity change imparted for the three scenarios in this study, parameterized over asteroid diameter. The reference diameter for each case is indicated with a black marker}
        \label{fig:flyby_dv}
    \end{center}
\end{figure}

For the parameters in this study, which will be described in subsequent sections, the $\Delta$v can be extremely small--on the order of 0.01 nm/s to 0.01 $\mu$m/s, as plotted in Figure~\ref{fig:flyby_dv}. The challenge for planetary defense flyby reconnaissance is to measure this small $\Delta$v with a meaningful confidence.

\section{Concept of Operations}\label{sec:conops}
We consider a host spacecraft that releases a maneuverable CubeSat-scale test-mass (TM) spacecraft 12 days prior to close approach. Unlike prior concepts, we are here proposing that the test-mass remains distant from the asteroid, while the more capable host spacecraft targets a low-altitude close approach. The Double Asteroid Redirection Test (DART) mission successfully demonstrated that a spacecraft could deploy a CubeSat without compromising its ability to target a precise encounter with a small asteroid using a combination of ground-computed human-in-the-loop navigation and on-board autonomous navigation \cite{Daly2023}. Our test-mass need only conduct ground-commanded maneuvers to account for updates to the asteroid's relative position and to reduce unwanted drift relative to the host spacecraft. (We envision that commands can be uplinked to the test-mass by using the host spacecraft as a relay. This eliminates the need for the test-mass to close a link with Earth). In the approach targeting sequences, the test-mass only conducts two maneuvers, while the host conducts either two or three depending on the approach speed (higher speed flybys benefit from an additional targeting maneuver). The host's final maneuver is set to be 12 hours prior to close approach. The common parameters for each scenario are given in Table~\ref{tab:common_param}.

\begin{figure}
    \begin{center}
    \includegraphics[width=0.5\columnwidth]{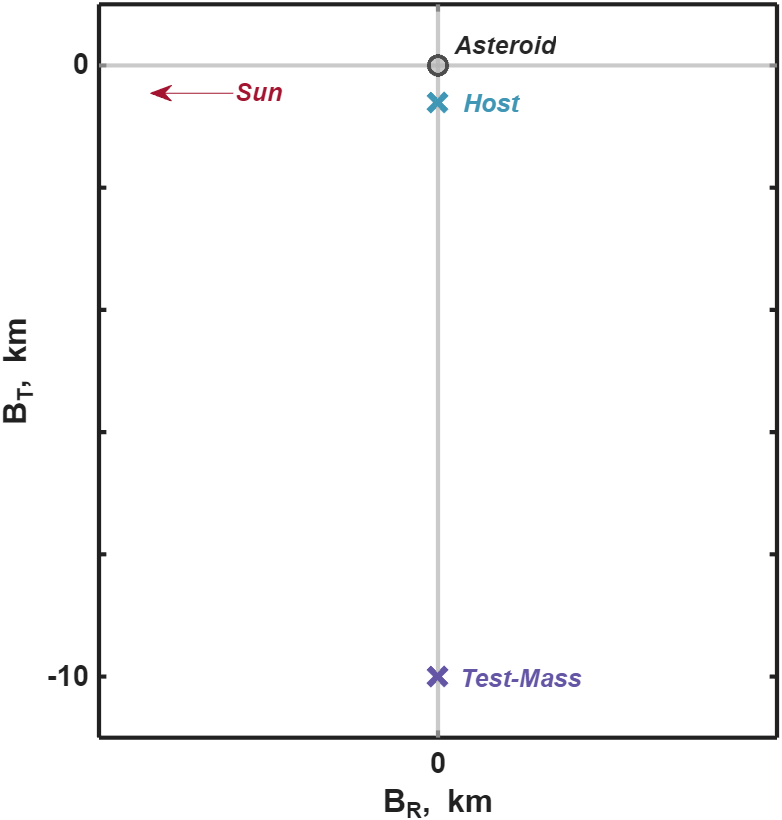}
    \caption{\small{Nominal B-plane targets for a 0.3 km diameter asteroid. The velocity vectors of the two spacecraft are perpendicular to the plane show (\textit{i.e.}, into the page).}}
    \label{fig:nom_b_plane_tgts}
    \end{center}
\end{figure}

Figure~\ref{fig:nom_b_plane_tgts} illustrates the nominal B-plane targets for the host and test-mass. The host is targeted to pass at 4 body radii from the center of the asteroid. This gives it an effective altitude of 3 body radii from the asteroid's spherical equivalent surface. Given expected uncertainties in the asteroid's size and aspect ratio, this is likely the lowest safe altitude a spacecraft might reasonably target. The location of the host's targeted B-plane target is oriented relative to the direction of the Sun. We selected a location that orients the asteroid's gravity to be orthogonal to solar radiation pressure, which is discussed further below.

The test-mass's close approach distance is 10 km, which means that it is effectively unperturbed by the asteroid. It serves as a (relatively) distant ballistic reference that the host can measure its motion against. It is oriented in the same direction as the host so that the asteroid's perturbation is oriented in the range and range-rate direction.

The asteroid's diameter is a parameter we vary, with the goal of assessing the mass measurement performance by target size for the three different scenarios. Table~\ref{tab:ast_diam} lists the asteroid sizes we consider, with the host's targeted distance for each. The asteroid is modeled to have a fixed density of 2,000 kg/m$^3$ (2 g/cm$^3$). The smallest size of 50 m is consistent with the Decadal recommendation for a future planetary defense mission \cite{NatAcad2022}, and is the smallest size for which US and international protocols would currently consider mounting an in-space response to \cite{Smpag2017,Nitep2021}. The 140 m size is consistent with the definition of a PHA, where this corresponds to an object with an absolute magnitude of 22 and an albedo of 0.14\footnote[2]{https://cneos.jpl.nasa.gov}. The absolute magnitude is used in this study to estimate the date that optical navigation measurements become available. 

The host and test-mass trajectories are meant to be as ``quiet'' as possible. To this end, they are envisioned to use reaction wheels for attitude control. The test-mass may outgas after its deployment, though we do not consider that possibility in this analysis.

The predominant uncertain perturbation is solar radiation pressure (SRP). We solve for a single scale-factor constant for each spacecraft, which approximates the spacecraft as a sphere. In ODTK, this is accomplished by using first-order Gauss Markov parameters with half-lives of 1 year, making them effectively constants. Additionally, we consider a strategy to reduce the SRP disturbance with the following: 1.) the B-plane targets are selected such that the nominal SRP perturbation acts orthogonal to the asteroid's gravity; 2.) the test-mass does not have to conduct any attitude motion outside of its short maneuvers; and 3.) both the host and test-mass have star trackers so that we can accurately reconstruct the attitude after-the-fact to support more elaborate N-plate models. Dedicated calibration campaigns pre-launch and earlier in cruise can help refine the SRP model. The simulations start with the host and test-mass having SRP \textit{a-priori} uncertainties of 5\% 3$\sigma$ and 20\% 3$\sigma$ respectively.

\begin{table}[!ht]
    \centering
    \caption{Common Simulation Parameters}
    \label{tab:common_param}
    \begin{tabular}{lr}
        \toprule
        Parameter & Value\\
        \midrule
        Host Mass              & 500.0 kg\\
        Host SRP Area          & 5.5 m$^2$\\
        Host SRP Nominal Cr    & 1.0 \\
        Host SRP Cr \textit{a-priori} 1$\sigma$    & 1.667\% \\
        Host Maneuver Magnitude Error 1$\sigma$  & 0.5\%\\
        Host Maneuver Pointing Error 1$\sigma$  & 0.5$^\circ$\\
        \midrule
        TM Separation   & 10 cm/s \\
        TM Separation 1$\sigma$ (per axis) & 5 cm/s \\
        TM Mass         & 12.0 kg\\
        TM SRP Area          & 0.2 m$^2$\\
        TM SRP Nominal Cr    & 1.0 \\
        TM SRP Cr \textit{a-priori} 1$\sigma$    & 6.667\% \\
        TM Maneuver Magnitude Error 1$\sigma$  & 0.5\%\\
        TM Maneuver Pointing Error 1$\sigma$  & 0.5$^\circ$ \\
        \bottomrule
    \end{tabular}
\end{table}

\begin{table}[!ht]
    \centering
    \caption{Simulated Asteroid Sizes}
    \label{tab:ast_diam}
    \begin{tabular}{crcc}
        \toprule
        Sph. Equiv.    &          & Host Flyby & Abs. \\
        Diam., km      & Mass, kg & Radius, km & Mag  \\
        \midrule
        0.050   & 1.31$\times10^8$  & 0.100 & 24.3\\
        0.100   & 1.05$\times10^9$  & 0.200 & 22.8\\
        0.140   & 2.87$\times10^9$  & 0.280 & 22.0\\
        0.200   & 8.38$\times10^9$  & 0.400 & 21.3\\
        0.300   & 2.83$\times10^{10}$ & 0.600 & 20.4\\
        0.400   & 6.70$\times10^{10}$ & 0.800 & 19.8\\ 
        0.500   & 1.31$\times10^{11}$ & 1.000 & 19.3\\
        0.800   & 5.36$\times10^{11}$ & 1.600 & 18.2\\
        1.000   & 1.05$\times10^{12}$ & 2.000 & 17.8\\
        \bottomrule
\end{tabular}
\end{table}

\subsection{Modeling Approach}
The estimation process is constructed as though the flyby has already occurred and all data has been downlinked. The truth states of each simulation model perfect execution of maneuvers to achieve the desired B-plane targets and close approach epochs. The predominant shortcoming of this approach is that we have to determine after-the-fact if the maneuver designs were possible with the available knowledge at the time of their effective data cut-off. 

The simulated measurements are processed to solve for the following states: the host spacecraft position, velocity, solar radiation pressure correction, and impulsive maneuvers; the test-mass position, velocity, solar radiation pressure correction, and impulsive maneuvers; the asteroid's position, velocity, and mass; and range measurement biases. No additional parameters are considered and no process noise or stochastic accelerations are added.

Given that the mass measurement relies on a very accurate representation of the dynamics (accelerations and noise terms), we validated the simulations by using two independent orbit determination tools: the NASA JPL tool MONTE \cite{Evans2018} and the commercial tool Orbit Determination Tool Kit (ODTK) \cite{vallado2010}. Despite differences in their formulations, the tools produced consistent results for our reference scenarios, which gives us confidence in the realism of this study's results.

\subsection{Measurements} 
The host is tracked by the Deep Space Network (DSN), as is typical of prior flyby missions. We assume 3 passes per week during cruise, one pass per day in the final 10 days near close approach, and continuous coverage during the four days before and after close approach. Although there are periods of time when the spacecraft would have dual coverage, we only model single station coverage throughout the scenario. The ranging passes are assumed to use modern Pseudo Noise (PN) modulation in X or Ka bands. We also exclude Delta-Differential One-Way Ranging (Delta-DOR) measurements in favor of some mild conservatism.

The host spacecraft has a camera on-board, which is used for optical navigation, ``OpNav'', images of the asteroid. We model the camera with accuracy and noise characteristics similar to those of the DRACO instrument that flew on the DART mission \cite{Fletcher2018}. The optical measurements are modeled as inertial right ascension and declination pairs with a 0.2 (binned) pixel 1$\sigma$ centroid white noise. We assume a ``burst'' of 6 images every 4 hours, once the asteroid is brighter than a visual magnitude of 18, a value achievable with a $<$60 second integration time. In the $\pm4$ hours around close approach, if the asteroid is observable, the imaging cadence increases to an OpNav measurement every 10 minutes, to simulate nearly continuous pointing typical of encounter operations. 

The test-mass is deployed shortly after the first optical navigation images are downlinked, giving the team an opportunity to execute a large correction maneuver if necessary. We model the deployment as a fixed 10 cm/s separation with a large \textit{a-priori} uncertainty of 5 cm/s (1$\sigma$) about each axis. This deployment impulse acts on both spacecraft in opposite directions, proportional to their masses (9.77 cm/s to the test-mass and 0.23 cm/s to the Host). The 5 cm/s per-axis uncertainty is likewise divided between the two impulses.

The host is able to track the neighboring test-mass optically to derive right-ascension and declination angles, a measurement type we call ``OpGrav''. The OpGrav measurements are taken with the same camera used for OpNav. We model a burst of 6 OpGrav measurements after every OpNav measurement. This begins 1 hour after deployment and continues until the end of the simulation, with the exception of $\pm4$ hours around close approach when it is assumed that the spacecraft will only be imaging the asteroid.

The host also constructs two-way intersatellite range and Doppler measurements with the test-mass. Unlike typical ground-links, the short intersatellite distance enables accurate measurements without the need for high precision oscillators \cite[e.g., ][]{Molli2023, Dibenedetto2019}. We call the intersatellite RF range and Doppler measurements ``DopGrav''. These measurements are generated every minute continuously, starting an hour after the test-mass's deployment. In our simulations, we assume that the spacecraft can close this intersatellite link nearly continuously regardless of its attitude using low-gain antennas. This set of measurements (DSN, OpNav, OpGrav, and DopGrav) represent our baseline measurement techniques. In an effort to evaluate potential near-term technologies, we also evaluate a laser ranging instrument (LRI) used to provide high precision range measurements, and a high precision Doppler (HPD) instrument. 

The LRI is modeled after the Small All-Range Lidar (SALI) instrument \cite{Sun2021}, which can range to natural targets at $>$100km distance and can achieve at closer range 1 cm 1$\sigma$ range measurements \cite{Sun2024}. SALI is a swath-mapping laser altimeter recently developed to Technology Readiness Level (TRL) 6 under a NASA Maturation of Instruments for Solar System Exploration (MatISSE) project. SALI uses a laser pulse train modulated with a pseudo-noise code and onboard, real-time correlation processing \cite{Cremons2021}. The test-mass can be equipped with a corner cube retroreflector facing the host spacecraft, whose position relative to the center-of-mass would be corrected using the test-mass's star-tracker based attitude solution. The high optical gain of the corner cubes allows 1-cm ranging from up to $\sim$20 km. To enable nearly continuous measurements in the encounter attitude, the laser divergence and detector FOV can be widened, given the SNR is very favorable when ranging to a retroreflector at close range ($\sim$10km). The multi-pixel detector also provides a wider FOV without mechanism. If spacecraft pointing performance is not sufficient, a small fine-pointing mirror can easily be added to the current SALI design to meet the pointing requirements. The size, mass, and power of SALI are all commensurate with a SmallSat platform, especially with the high optical gain from a retroreflector which would remove the need for the two-stage laser amplifier and reduce the instrument power significantly.

The HPD is modeled after the analog radio described in the GIRO hardware \cite{park2025}. This low-power radio can achieve 0.1 $\mu$m/s 1$\sigma$ range-rate measurements when the host spacecraft is equipped with an ultrastable oscillator. This is modeled with a 60 second integration time. We also note that this precise measurement accuracy levies a requirement that the host and test-mass attitude control systems be capable of reconstructing the true 60 second time-averaged attitude history to an accuracy of better than roughly 0.1 urad/sec and 2 urad/sec respectively. This allows us to distinguish the true translational range-rate from attitude disturbances causing the antenna phase centers to move relative to their spacecrafts' center-of-mass.

Table~\ref{tab:meas_params} lists these measurement accuracies. 

\begin{table}
    \centering
    \caption{\small{Measurement Accuracy Models}}
    \label{tab:meas_params}
    \begin{tabular*}{0.615\columnwidth}{lr}
        \toprule
        Measurement & 1$\sigma$\\
        \midrule
        DSN Doppler Noise                   & 0.1 mm/s \\
        DSN Range Noise                     & 0.5 m  \\
        DSN Range Per-Pass Bias             & 1.5 m \\
        Optical Meas (RA, Dec) Noise        & 1 urad\\
        Intersatellite Baseline RF Range-Rate Noise  & 0.1 mm/s \\
        Intersatellite Baseline RF Range Noise       & 10.0 m \\
        Intersatellite Baseline RF Range Bias        & 1.5 m \\     
        \midrule
        Intersatellite LRI Range Noise      & 1 cm \\       
        Intersatellite LRI Range Bias       & 1.5 m \\      
        \midrule
        Intersatellite HPD Range-Rate Noise & 0.1 $\mu$m/s \\       
        \bottomrule
    \end{tabular*}
\end{table}

\section{Case 1: 2023 PDC}\label{sec:case1}
\subsection{Scenario Definition}
The first case we consider is the 2023 Planetary Defense Conference hypothetical threat\footnote[1]{https://cneos.jpl.nasa.gov/pd/cs/pdc23/}, which was named 2023 PDC. This is the slowest flyby case, with a speed of only 1.7 km/s. We formerly studied this case with fewer maneuvers, a less realistic operation measurement cadence, the test-mass targeting the low-altitude close approach, and different B-plane targets \cite{Bull2024}. 

\begin{figure*}[tbh!]
  \begin{tabular}{cc}
  \multicolumn{2}{c}{(a)} \\
  \multicolumn{2}{c}{\includegraphics[width=0.5\textwidth]{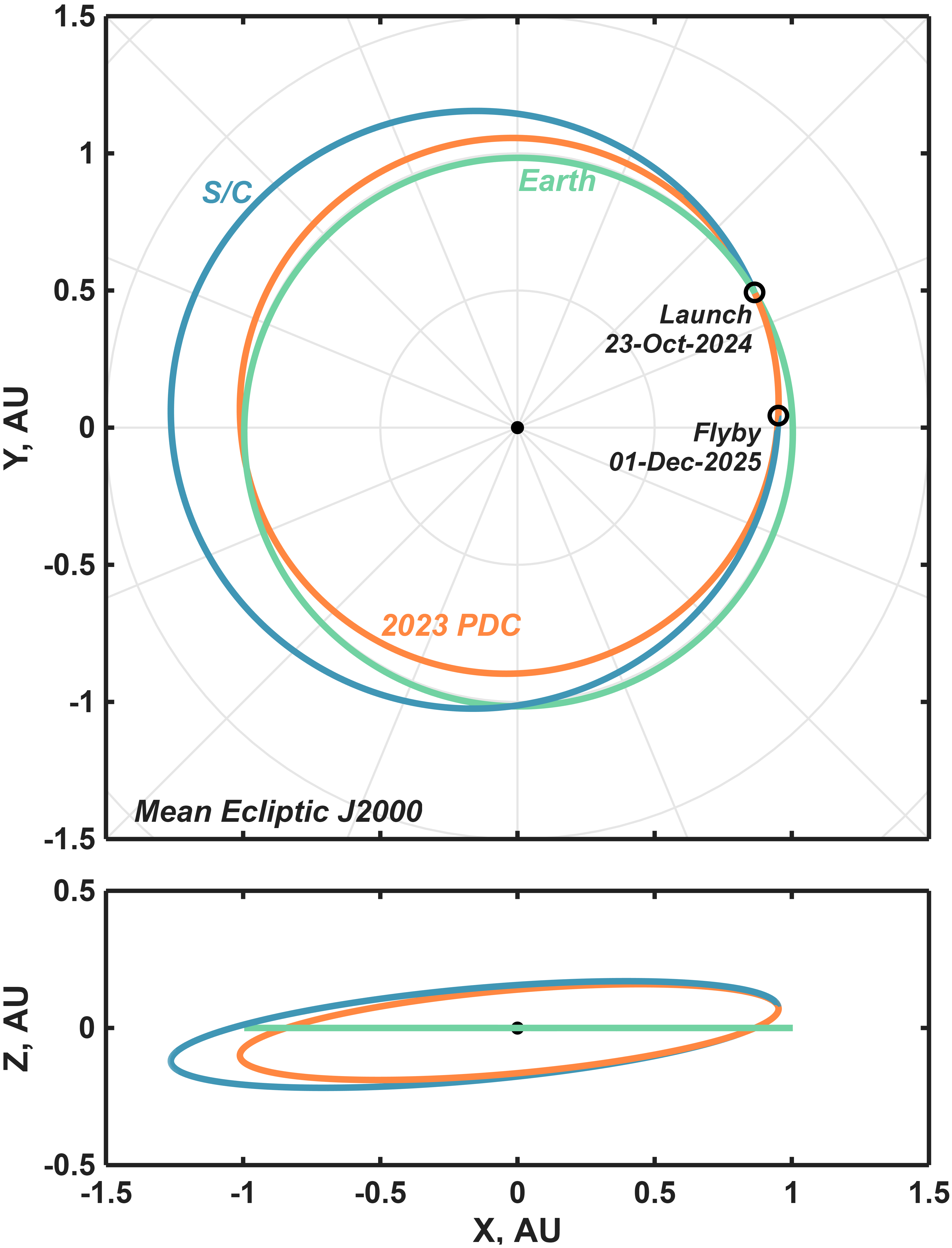}} \\
  \\
    (b) & (c) \\
  \includegraphics[width=0.425\textwidth]{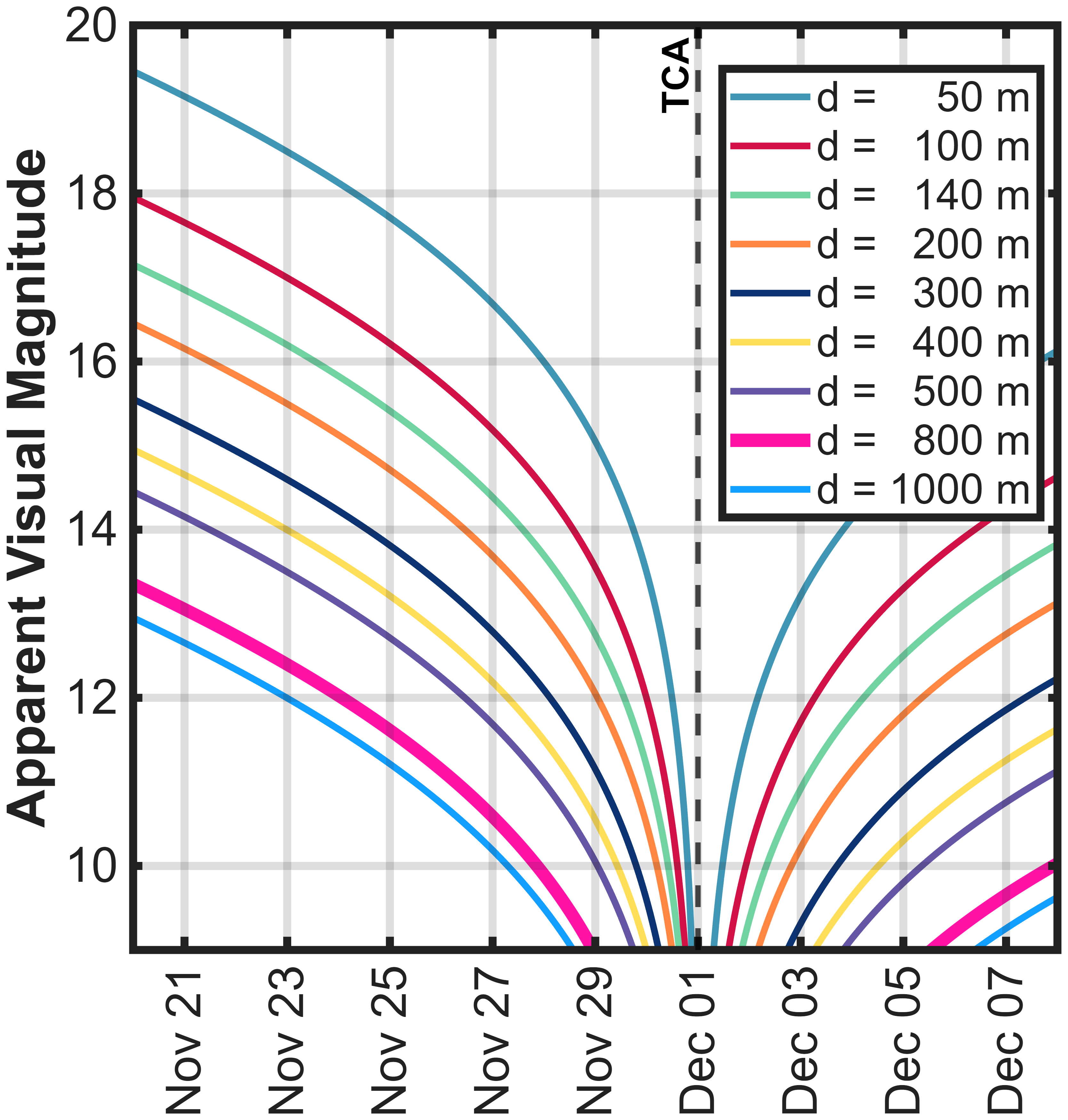} &
  \includegraphics[width=0.475\textwidth]{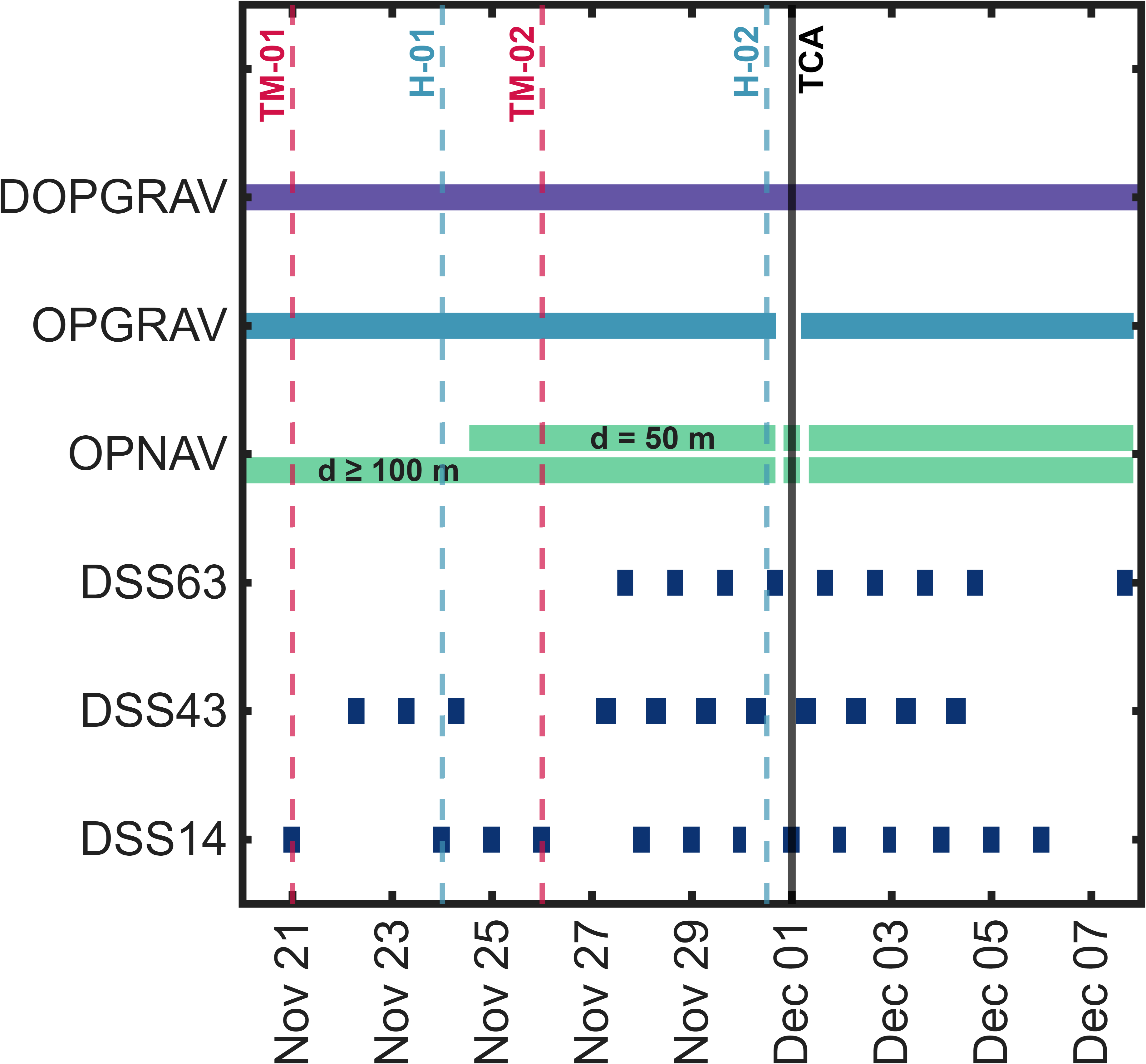} \\
  \\
  \end{tabular}
\caption{
Case 1. (a) Flyby trajectory to 2023 PDC. (b) Temporal evolution of the apparent magnitude of 2023 PDC from the spacecraft for varying asteroid diameters. The vertical line indicates the time of closest approach (TCA). (c) Measurement schedule for the encounter with 2023 PDC. The vertical dashed lines indicate the times of host (H) and test-mass (TM) maneuvers.
\label{fig:case_1_inputs}}
\end{figure*}

The interplanetary trajectory is shown in Figure~\ref{fig:case_1_inputs}a. The apparent magnitude for a range of diameters is given in Figure~\ref{fig:case_1_inputs}b. The nominal measurement schedule is illustrated in Figure~\ref{fig:case_1_inputs}c. DSS-14, DSS-43, and DSS-63 correspond to DSN antennas at Goldstone, Canberra, and Madrid respectively. The solar phase angle over the pre-encounter time period decreases from 120$^\circ$ to 111$^\circ$. The close approach occurs on 1 Dec 2025. The OpNav images start at either the beginning of the simulation (12 days prior to closest approach) or when the apparent magnitude of the asteroid reaches 18, which depends on asteroid diameter. In this case, this only affects the smallest asteroid size, 50 m, for which OpNav starts 6.5 days prior to close approach. 

The asteroid's uncertainty at the start of the scenario was provided by the PDC scenario developers. The host spacecraft's uncertainty was produced by simulating three weekly DSN passes for the duration of its cruise, with a pre-separation maneuver conducted on 17 Nov 2025, 14 days before the flyby. Separation occurs on 19 Nov 2025, 12 days before the flyby. Our encounter simulations then start immediately after the separation. 

Both the host and test-mass perform two post-separation maneuvers to target their respective B-plane crossings and ensure a simultaneous flyby epoch. The maneuver epochs are illustrated as vertical dashed lines in Figure~\ref{fig:case_1_inputs}c. The maneuvers are designed to represent a scenario where a 25 km B-plane error is corrected 7 days out, and a 2 km B-plane error is corrected 12 hours out. This corresponds to host maneuvers of approximately 5 and 2.5 cm/s. The test-mass's maneuvers are approximately 80 cm/s and 15 cm/s. The test-mass's higher maneuver magnitudes are required to cancel the drift and separation that accumulated between its separation and its first maneuver.

\subsection{Results}
\begin{figure*}[tbh!]
  \begin{tabular}{cc}
  (a) & (b) \\
  \includegraphics[width=0.475\textwidth]{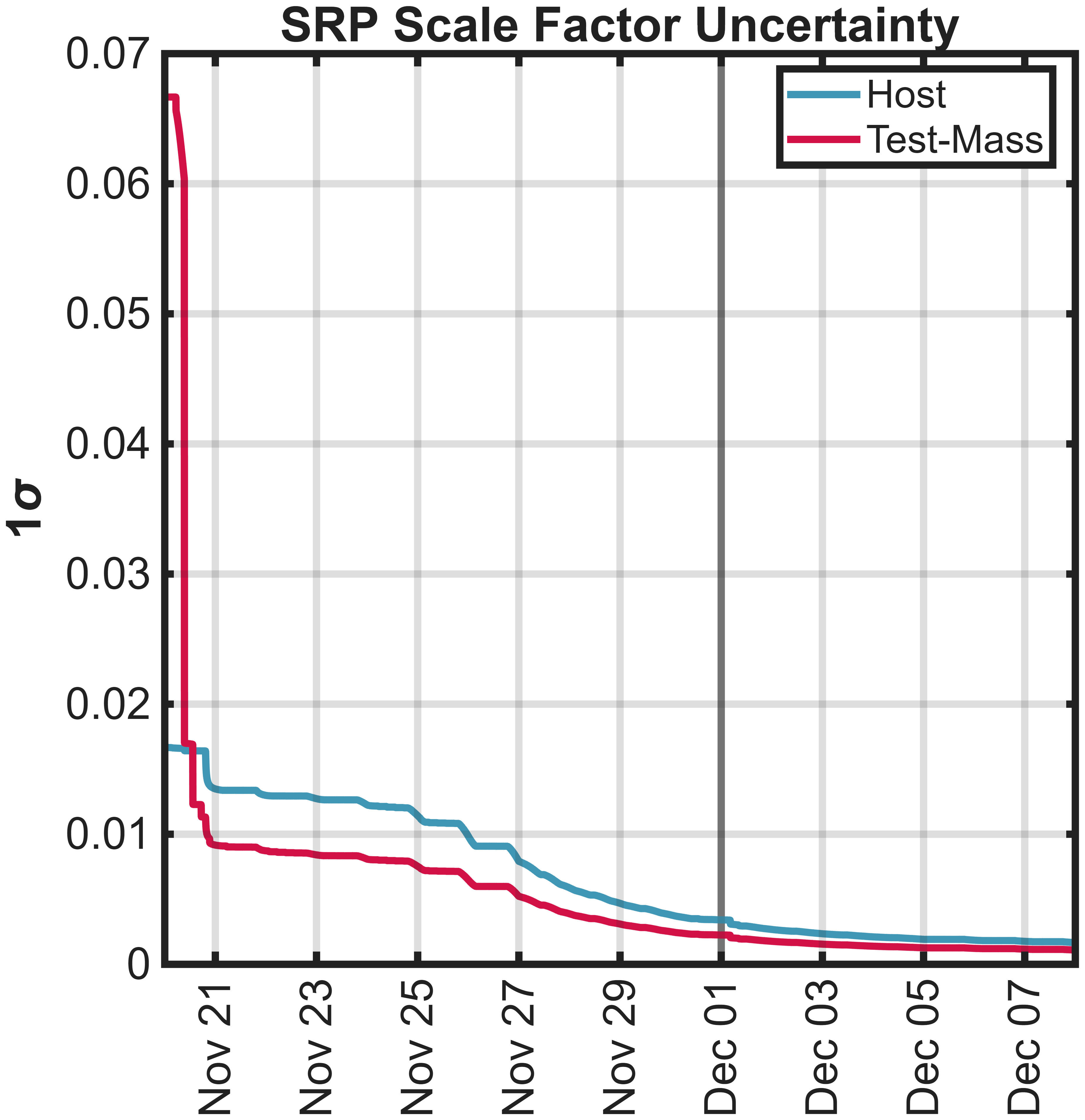} &
  \includegraphics[width=0.485\textwidth]{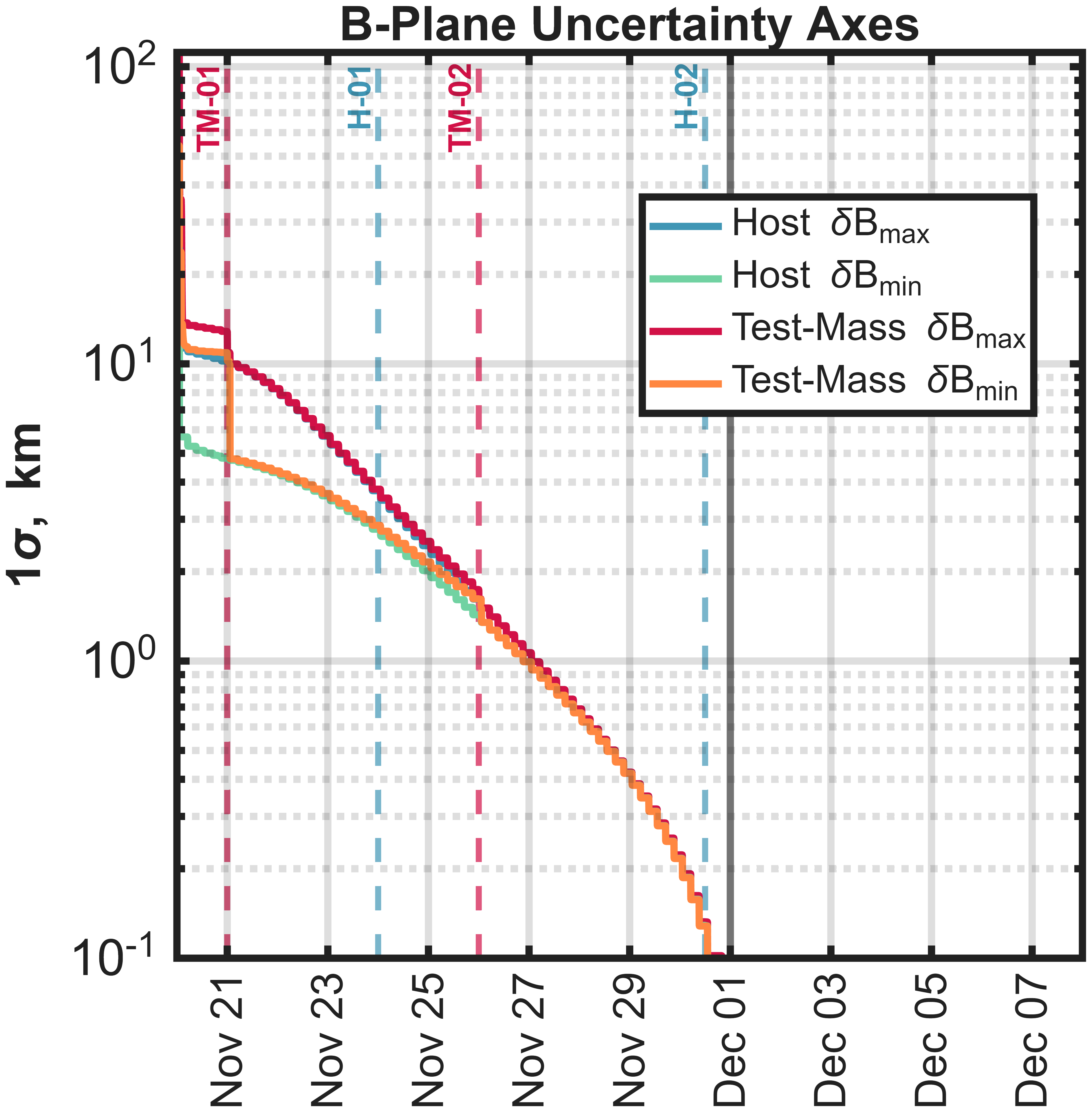} \\
  \\
    (c) \\
  \includegraphics[width=0.485\textwidth]{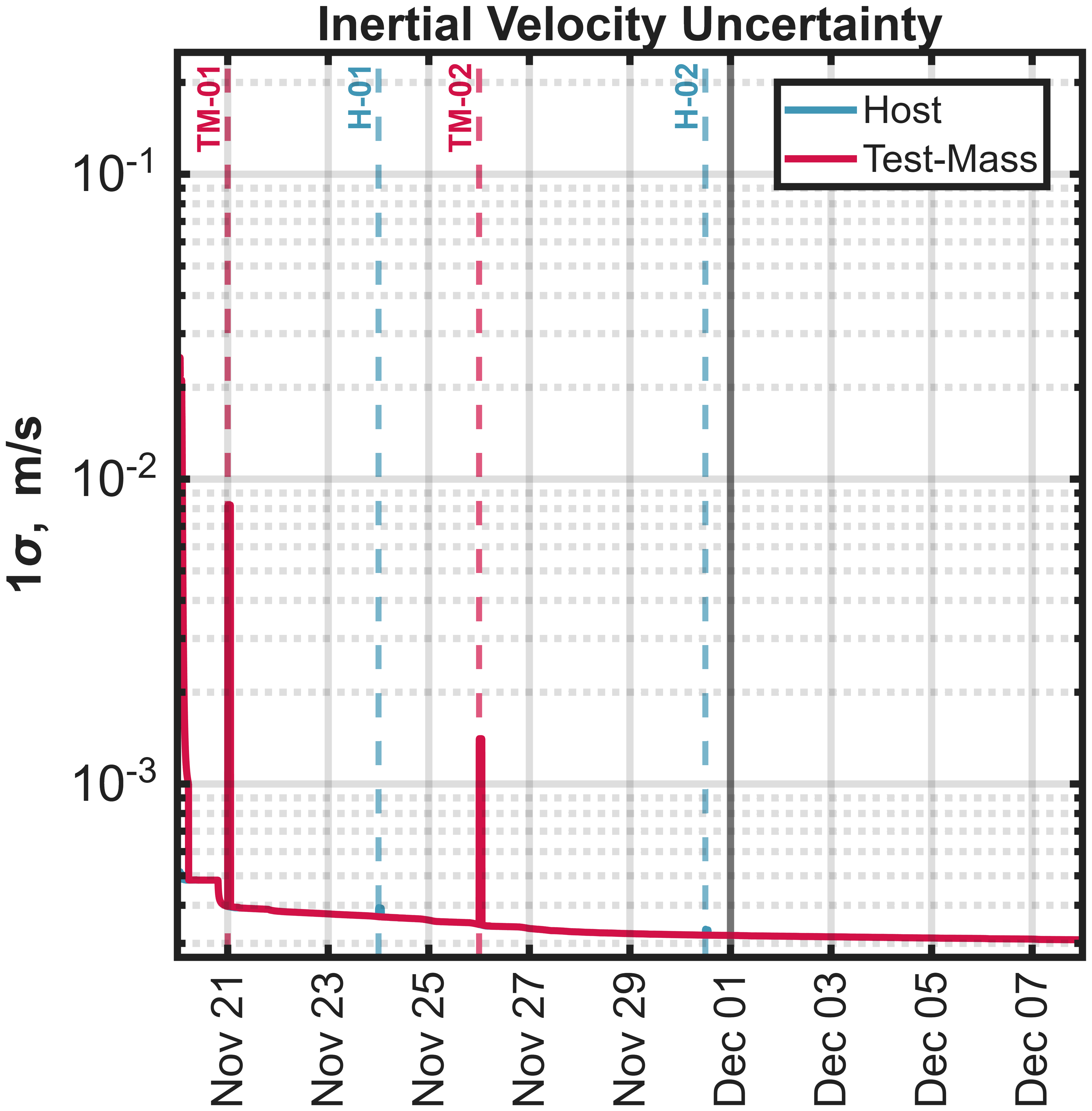} & \\
  \\
  \end{tabular}
\caption{
Case 1. Temporal evolution of spacecraft (a) SRP coefficient uncertainty, (b) B-plane uncertainty ellipse sizes for an encounter with an 800 m diameter asteroid, and (c) inertial velocity uncertainty.
\label{fig:case_1_outputs_1}}
\end{figure*}

\begin{figure*}[tbh!]
  \begin{tabular}{cc}
  (a) & (b) \\
  \includegraphics[width=0.500\textwidth]{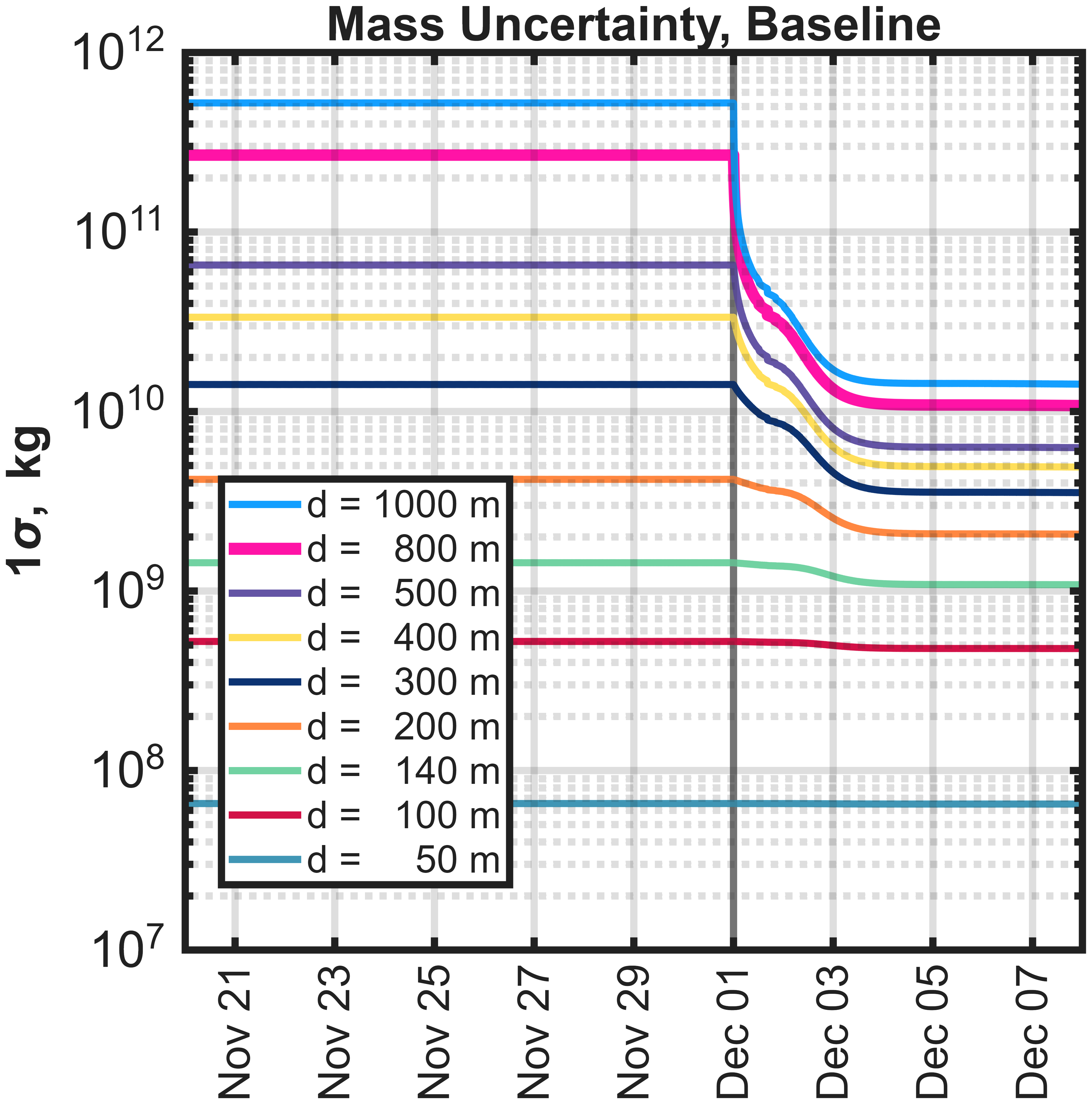} &
  \includegraphics[width=0.475\textwidth]{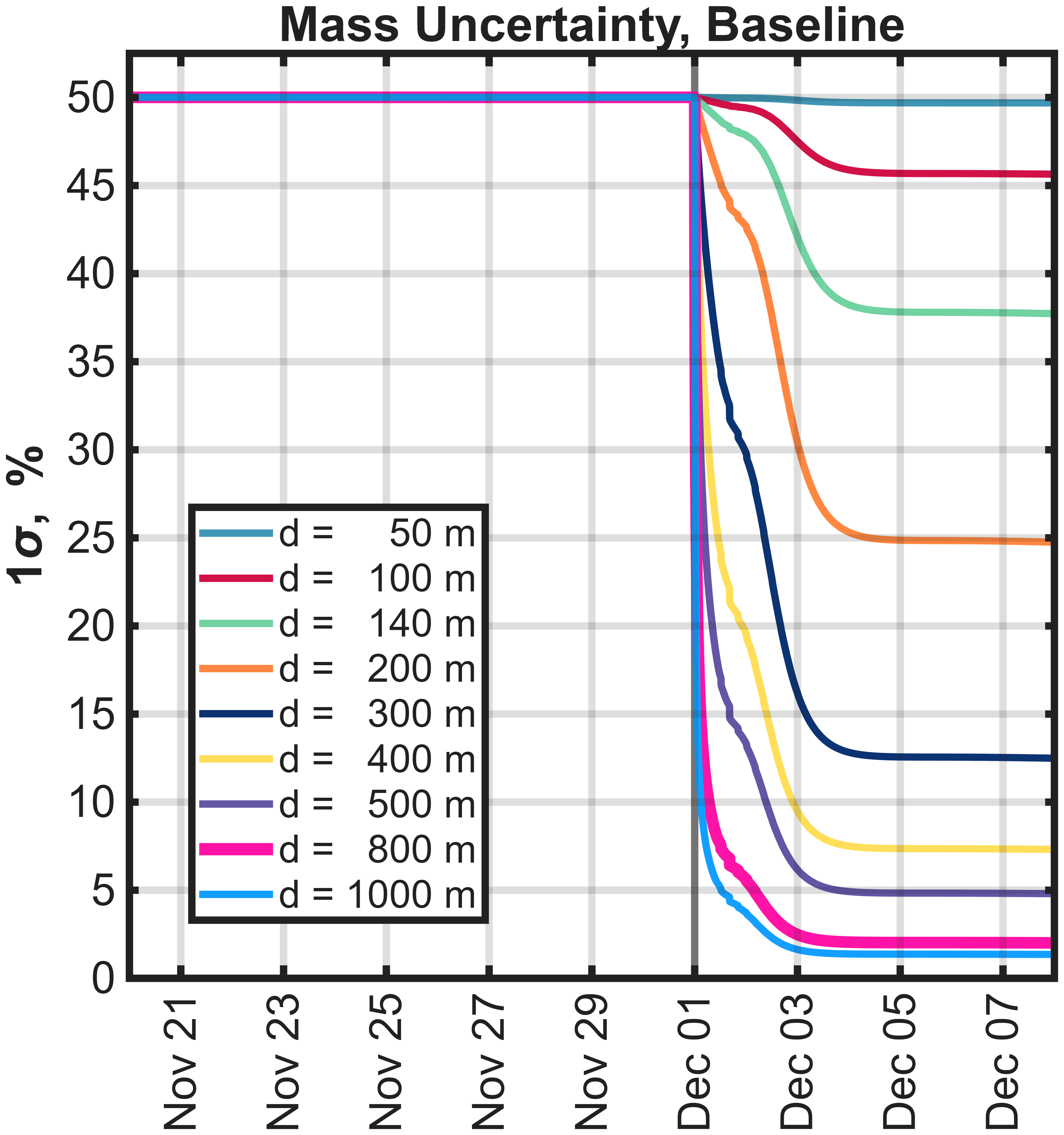} \\
  \\
    (c) & (d) \\
  \includegraphics[width=0.475\textwidth]{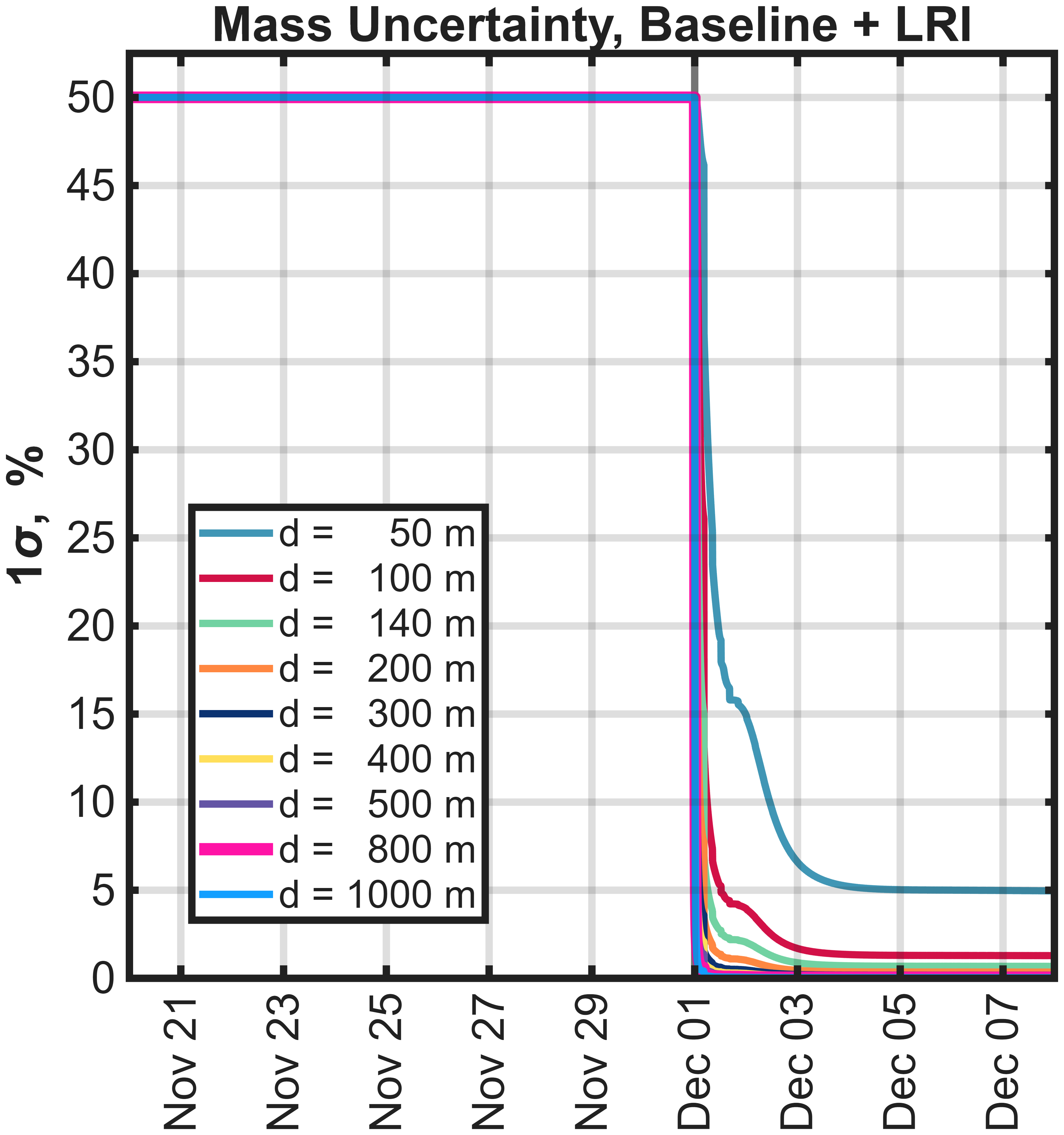} &
  \includegraphics[width=0.475\textwidth]{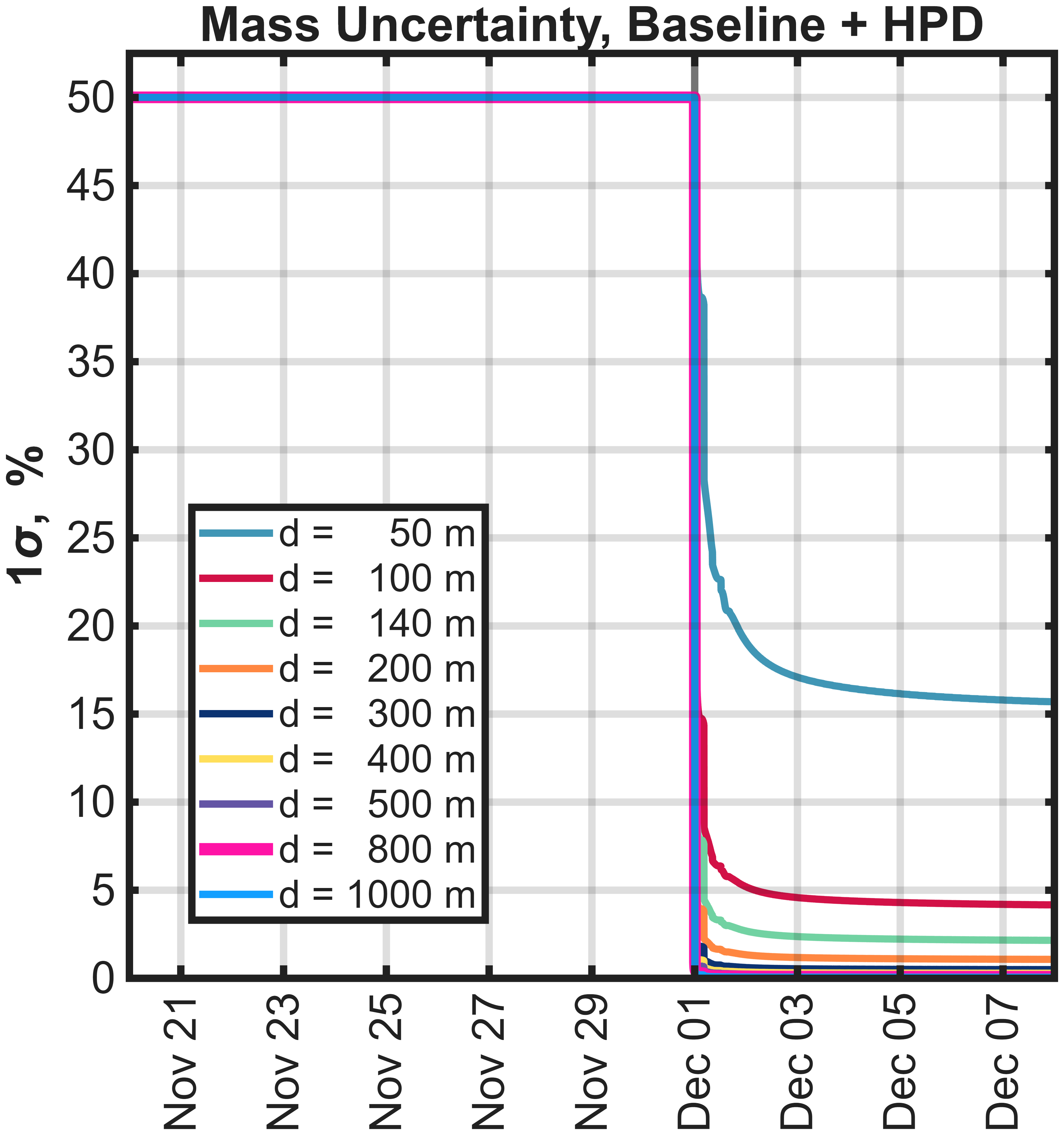} \\
  \\
  \end{tabular}
\caption{
Case 1. Temporal evolution of (a) asteroid mass uncertainty using baseline tracking measurements, (b) asteroid percent mass uncertainty using baseline tracking measurements, (c) asteroid percent mass uncertainty baseline tracking measurements augmented with LRI, and (d) asteroid percent mass uncertainty using baseline tracking measurements augmented with HPD.
\label{fig:case_1_outputs_2}}
\end{figure*}

\begin{figure}
    \centering
    \includegraphics[width=0.6\textwidth]{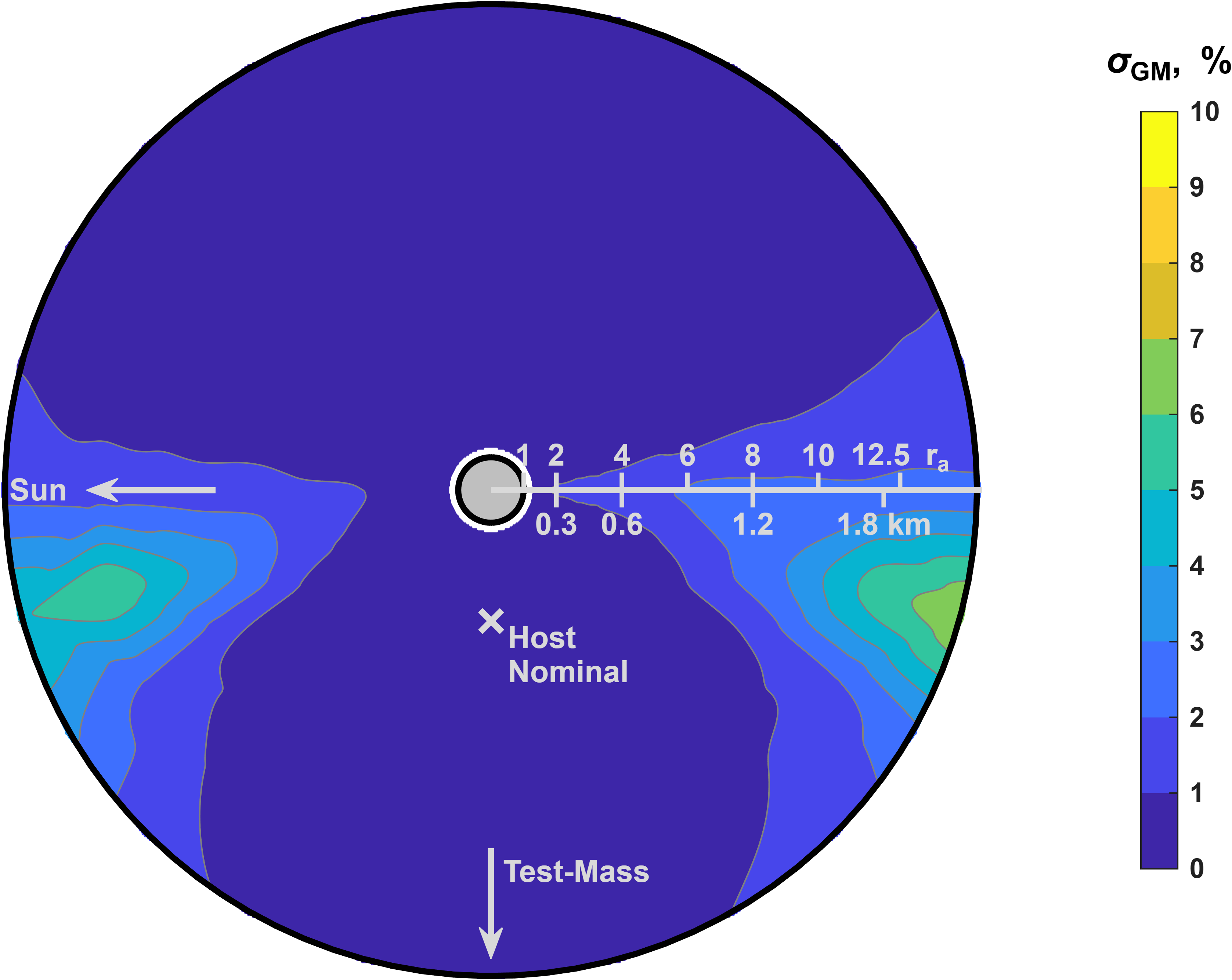}
    \captionof{figure}{Case 1. Final asteroid mass uncertainty as a function of host B-plane flyby location for a 300 m diameter asteroid with LRI augmented measurements}
    \label{fig:2023_PDC_bpln}
\end{figure}

The results of the covariance analysis are shown in Figures~\ref{fig:case_1_outputs_1}-\ref{fig:2023_PDC_bpln}. The figures include the temporal evolution of the $1\sigma$ uncertainties for a specified set of states. Figure~\ref{fig:case_1_outputs_1}a shows the evolution of the host and test-mass solar radiation pressure coefficients. The test-mass's SRP scale factor uncertainty drops quickly with intersatellite tracking data until it roughly matches the host’s SRP uncertainty. Over the duration of the simulation the pair’s SRP uncertainties remain highly correlated, settling to roughly 0.1\% by the end of the simulation. Figure~\ref{fig:case_1_outputs_1}b shows the B-plane 1$\sigma$ uncertainty ellipse semimajor ($\delta B_{max}$) and semiminor ($\delta B_{min}$) axes for both the Host and the test-mass. This represents the knowledge of the spacecraft relative to the asteroid. The B-plane sizes are identical for both spacecraft, which meets expectations that the test-mass's state uncertainty relative to the host is small compared to the uncertainty of the host relative to the asteroid. The maximum B-plane uncertainty at the time of the final host maneuver is roughly $\pm$130 m 1$\sigma$. Ideally, we would want the 3$\sigma$ uncertainty to be less than the selected flyby altitude, which is 3 body radii. This suggests that the computed B-plane knowledge is sufficient for targeting asteroids as small as 130 m in radius (260 m in diameter). After the flyby, the reconstructed B-plane knowledge is roughly 10 m 1$\sigma$. Figure~\ref{fig:case_1_outputs_1}c shows the host's and test-mass's inertial velocity uncertainties, with ``spikes'' sometimes apparent at the times of maneuvers. The lines for the host and test-mass overlap, indicating that their velocity uncertainties are otherwise nearly identical. The rapid decrease in uncertainty indicates that the maneuvers are very observable given immediate post-maneuver tracking. Figure~\ref{fig:case_1_outputs_2}a-b show the asteroid's mass uncertainty in absolute units (kg) and as a percent for the baseline measurement case. The uncertainty shows a characteristic drop at the epoch of the flyby, followed by asymptotic decrease to a final value approximately 3 days after close approach. Setting 25\% 1$\sigma$ as a meaningful metric, where our initial uncertainty is halved, we find that 200 m diameter asteroids are the smallest observable size in this scenario.

Figures~\ref{fig:case_1_outputs_2}c-d show the improvement available if the host is equipped with the laser ranging instrument (LRI) or high precision Doppler instrument (HPD). With either of these near-term instruments, the 50 m asteroid's mass is observable. Recalling that the computed B-plane targeting at 12 hours is not consistent with asteroid sizes below 140 m, this simulation indicates what is possible if a low flyby can be achieved in spite of this. To better characterize the expected performance in the presence of targeting errors, we produced a mapping of the computed mass measurement accuracy as a function of host flyby B-plane target location. Figure~\ref{fig:2023_PDC_bpln} shows the results for a 300 m diameter asteroid with the baseline tracking setup. This particular case has sufficient margin that it permits large errors of many body radii in either the radial or tangential directions, while still delivering better than 25\% final mass uncertainty.

\section{Case 2: 2024 PDC25}\label{sec:case2}
\subsection{Scenario Definition}
\begin{figure*}[tbh!]
  \begin{tabular}{cc}
  \multicolumn{2}{c}{(a)} \\
  \multicolumn{2}{c}{\includegraphics[width=0.5\textwidth]{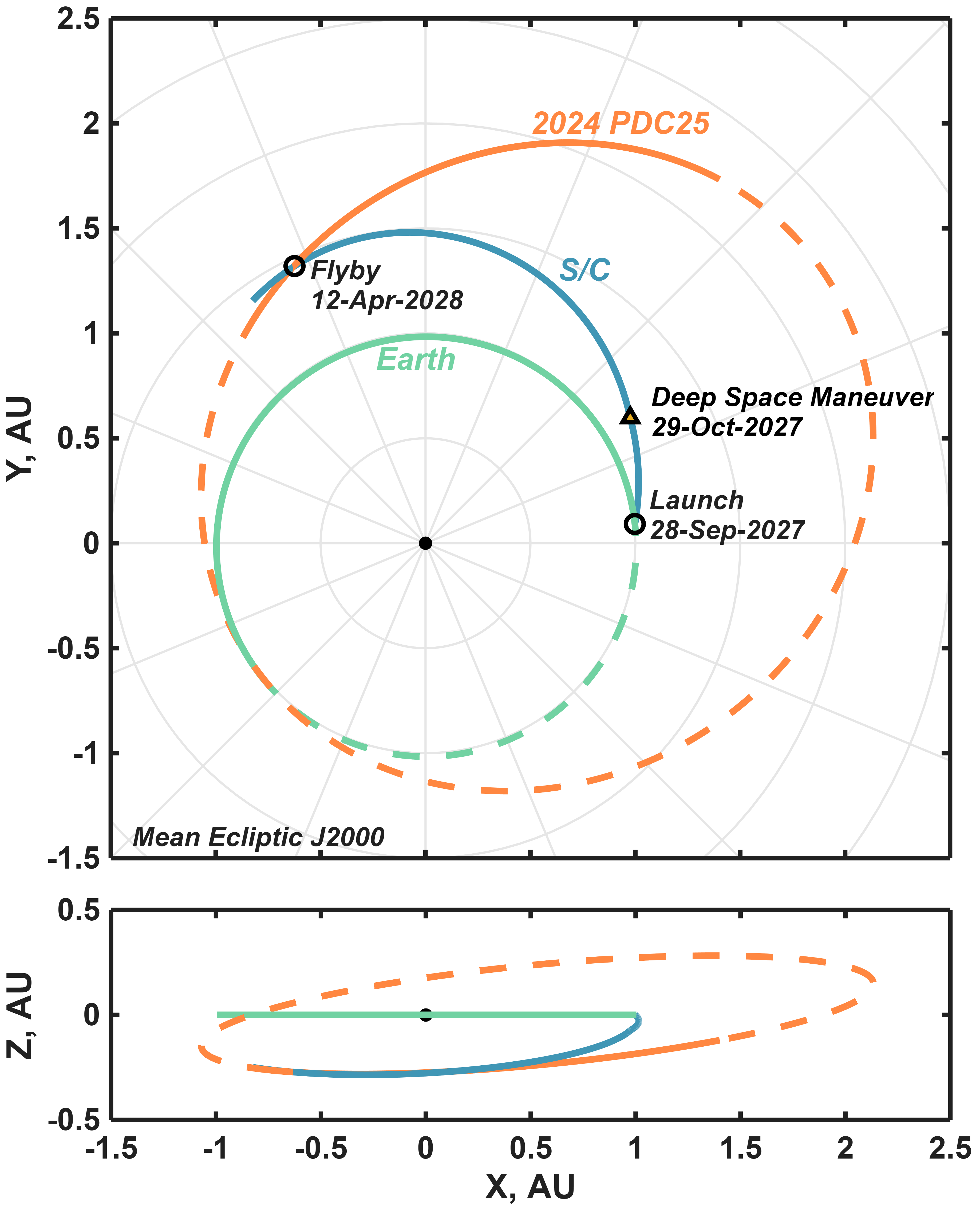}} \\
  \\
    (b) & (c) \\
  \includegraphics[width=0.425\textwidth]{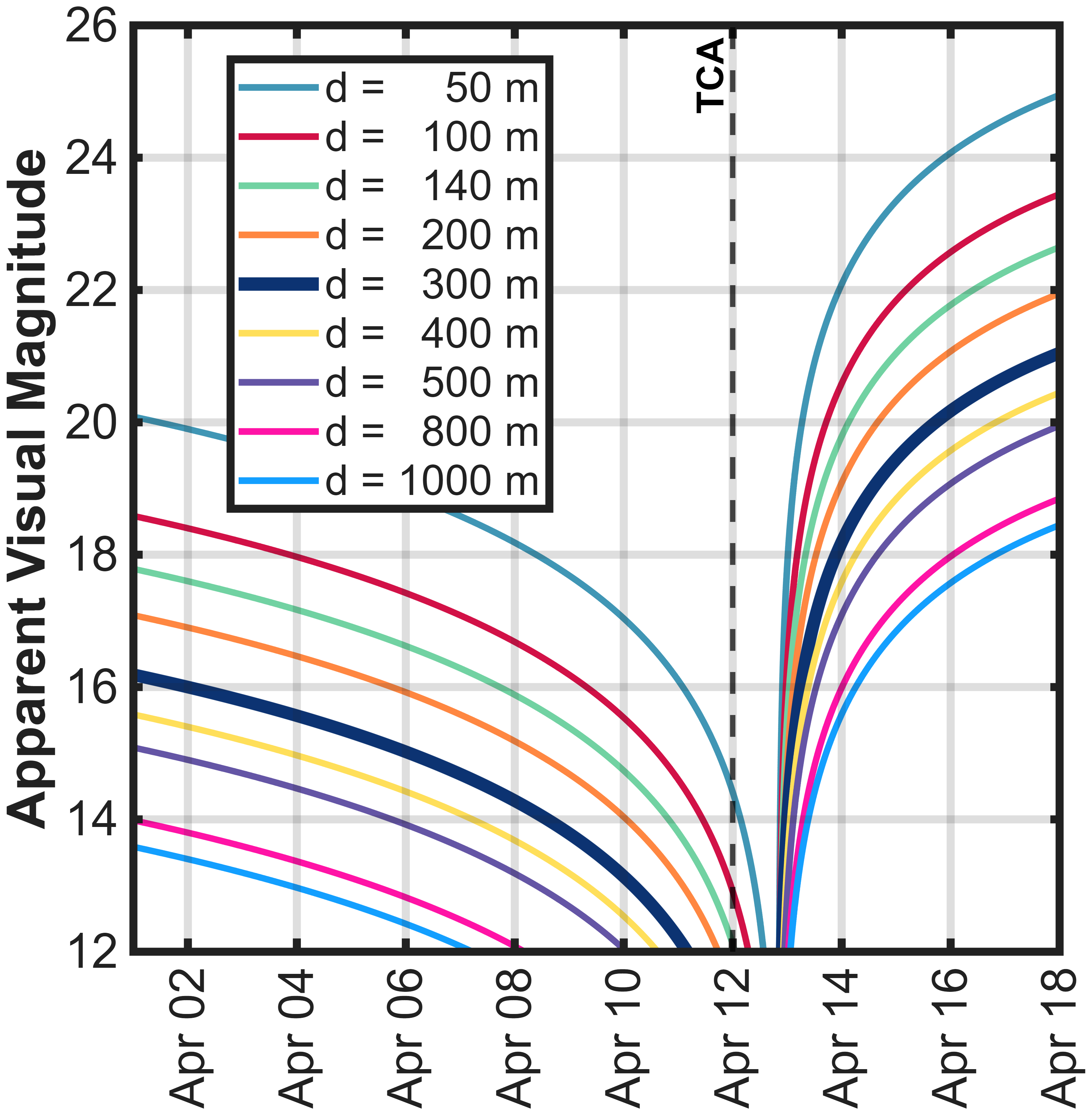} &
  \includegraphics[width=0.475\textwidth]{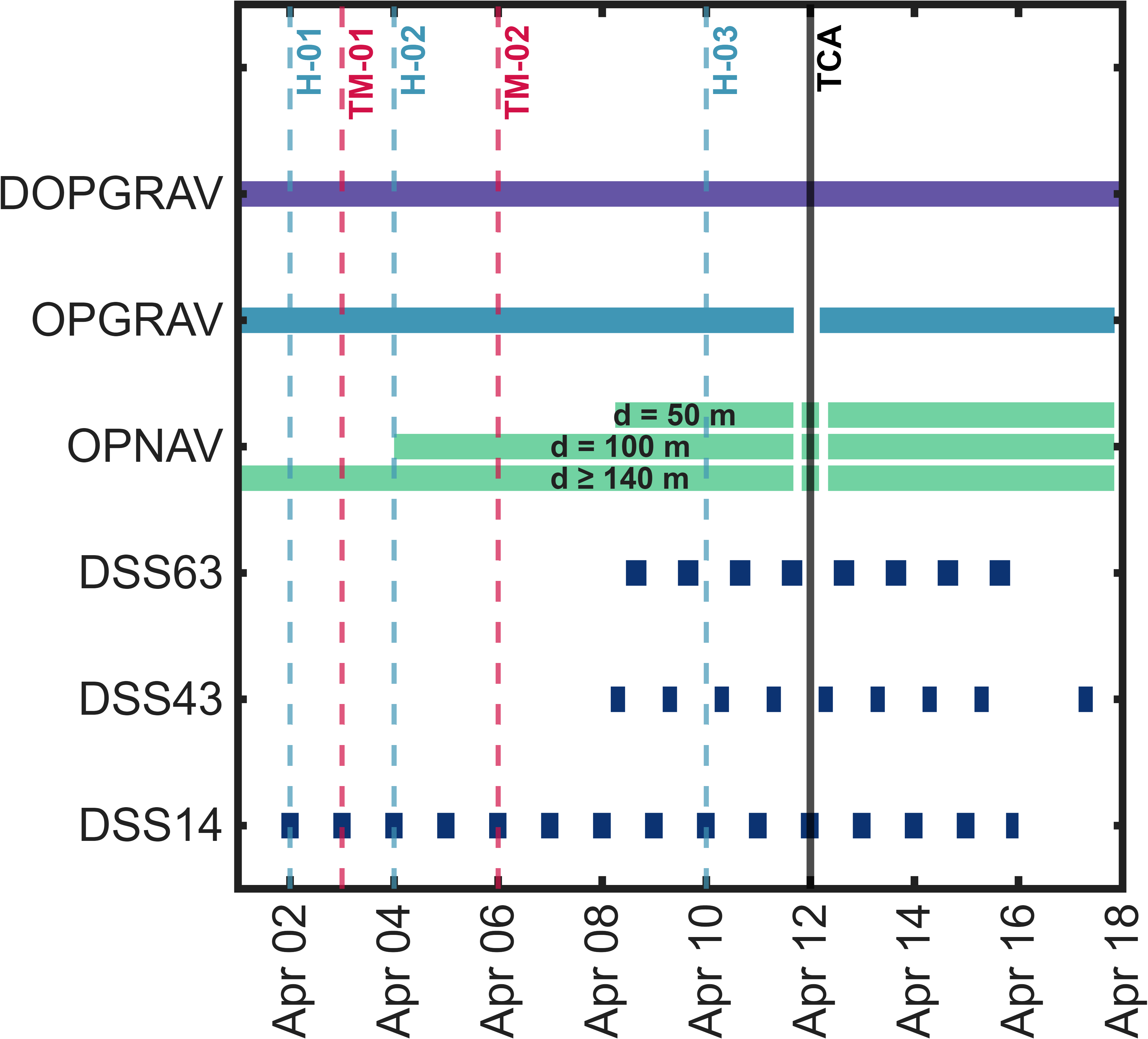} \\
  \\
  \end{tabular}
\caption{
Case 2. (a) Flyby trajectory to 2024 PDC25. (b) Temporal evolution of the apparent magnitude of 2024 PDC25 from the spacecraft for varying asteroid diameters. The vertical line indicates the time of closest approach (TCA). (c) Measurement schedule for the encounter with 2024 PDC25. The vertical dashed lines indicate the times of host (H) and test-mass (TM) maneuvers.
\label{fig:case_2_inputs}}
\end{figure*}

The second case we consider is the 2025 Planetary Defense Conference hypothetical threat\footnote[2]{https://cneos.jpl.nasa.gov/pd/cs/pdc25/}, which was named 2024 PDC25. This is the medium flyby-speed case, with a speed of 8.1 km/s. This scenario's reference asteroid diameter is 150 m. The interplanetary trajectory is shown in Figure~\ref{fig:case_2_inputs}a. The spacecraft trajectory includes a large deep space maneuver of 2 km/s one month after launch. 

 The apparent magnitude and the nominal measurement schedule are shown in Figures~\ref{fig:case_2_inputs}b \& \ref{fig:case_2_inputs}c. As before, the OpNav availability varies for each case, depending on the asteroid's simulated diameter. The solar phase angle over the pre-encounter time period increases from 24$^\circ$ to 30$^\circ$. The close approach occurs on 12 Apr 2028.

As in Case 1, the asteroid's uncertainty at the start of the scenario was provided by the PDC scenario developers and the host spacecraft's uncertainty was produced by simulating three weekly DSN passes for the duration of its cruise. Separation occurs on 01 Apr 2028, 12 days before the flyby. The encounter simulations then start immediately after the separation. 

The host performs three post-separations maneuvers, and the test-mass performs two maneuvers. The maneuver epochs are illustrated as vertical dashed lines in Figure~\ref{fig:case_2_inputs}c. We found that the additional host maneuver enabled a scenario where the host and test-mass are separating more slowly after the flyby.

\subsection{Results}
\begin{figure*}[tbh!]
  \begin{tabular}{cc}
  (a) & (b) \\
  \includegraphics[width=0.475\textwidth]{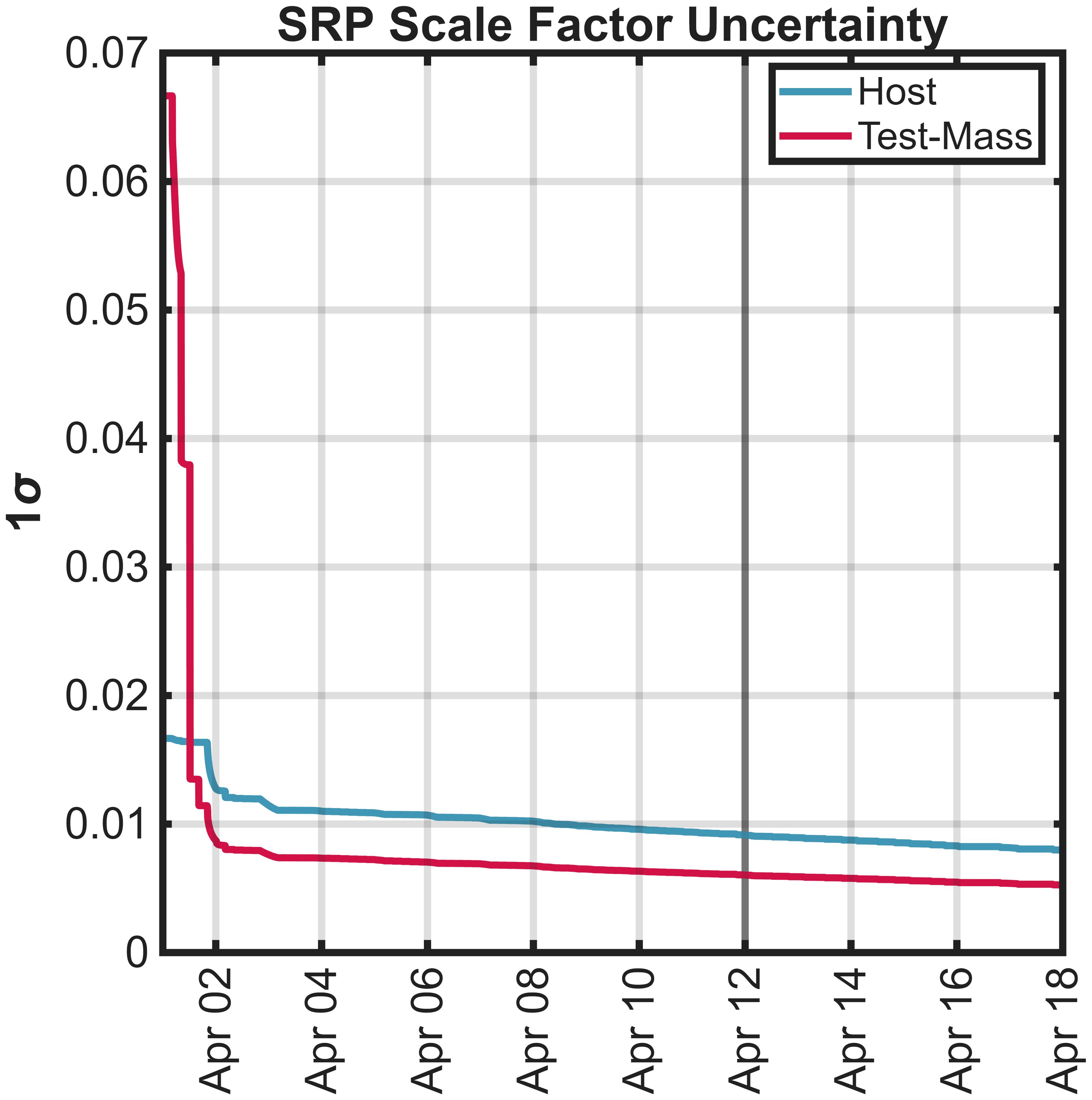} &
  \includegraphics[width=0.485\textwidth]{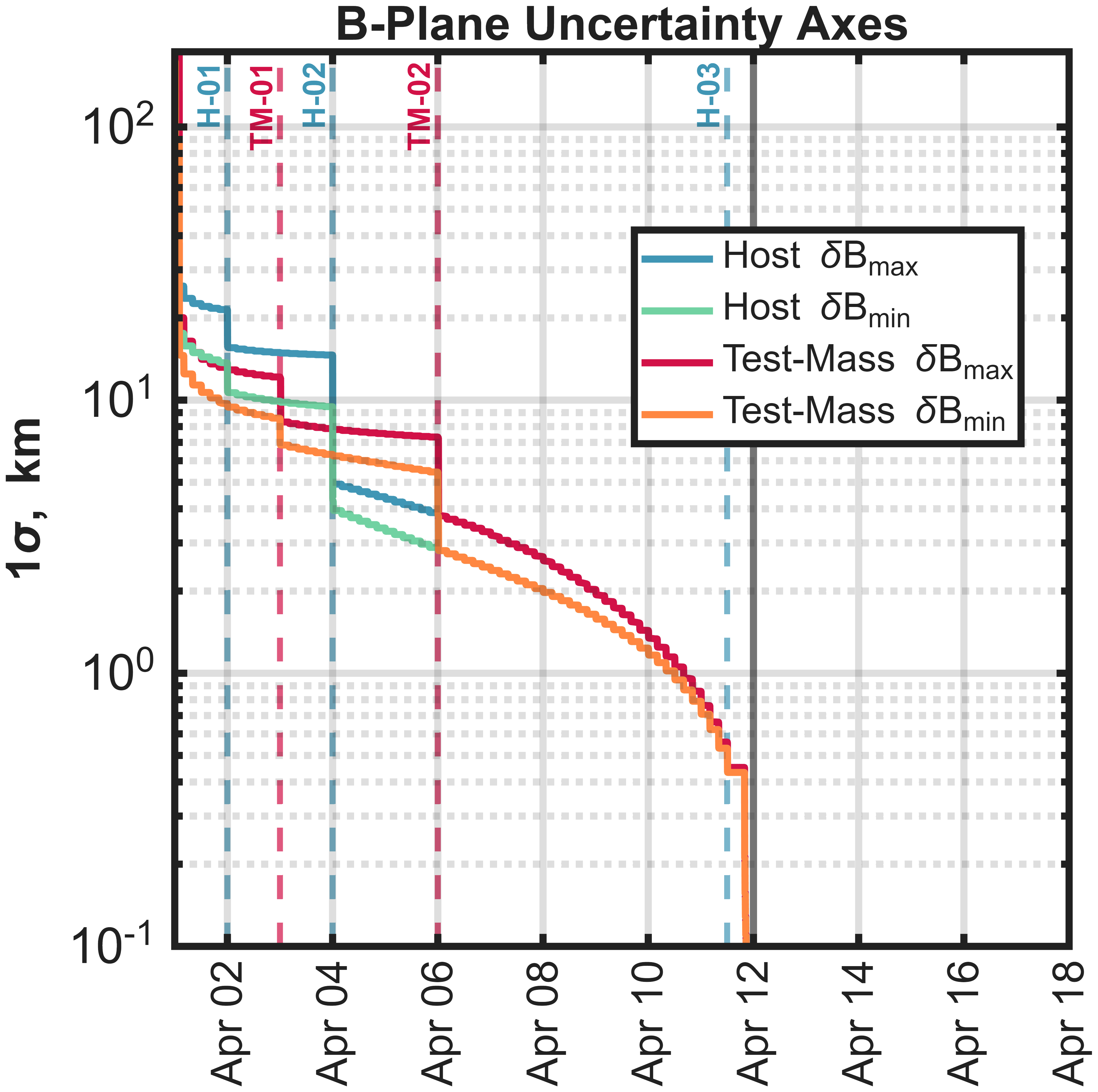} \\
  \\
    (c) \\
  \includegraphics[width=0.485\textwidth]{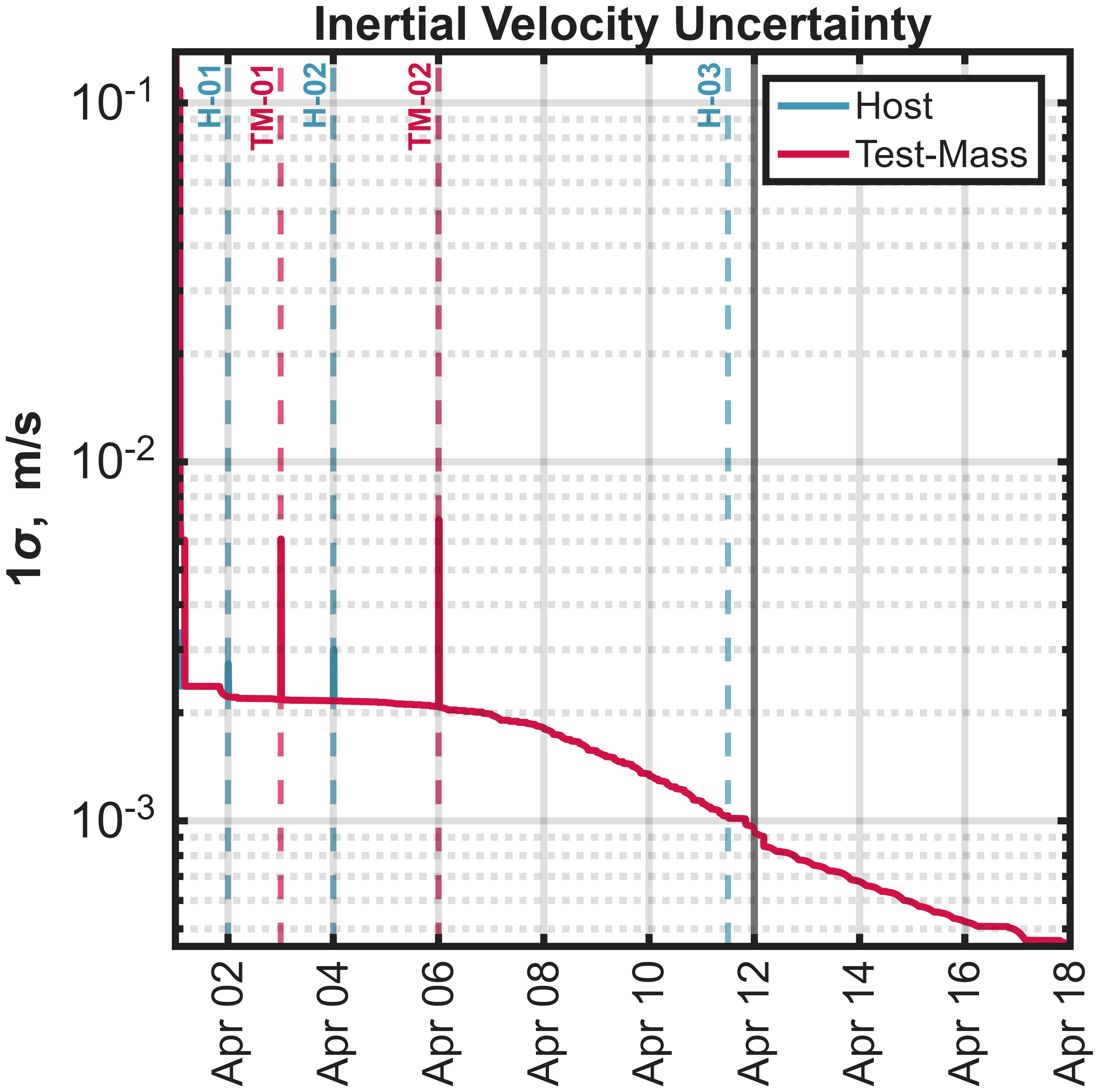} & \\
  \\
  \end{tabular}
\caption{
Case 2. Temporal evolution of spacecraft (a) SRP coefficient uncertainty, (b) B-plane uncertainty ellipse sizes for an encounter with an 300 m diameter asteroid, and (c) inertial velocity uncertainty.
\label{fig:case_2_outputs_1}}
\end{figure*}

\begin{figure*}[tbh!]
  \begin{tabular}{cc}
  (a) & (b) \\
  \includegraphics[width=0.500\textwidth]{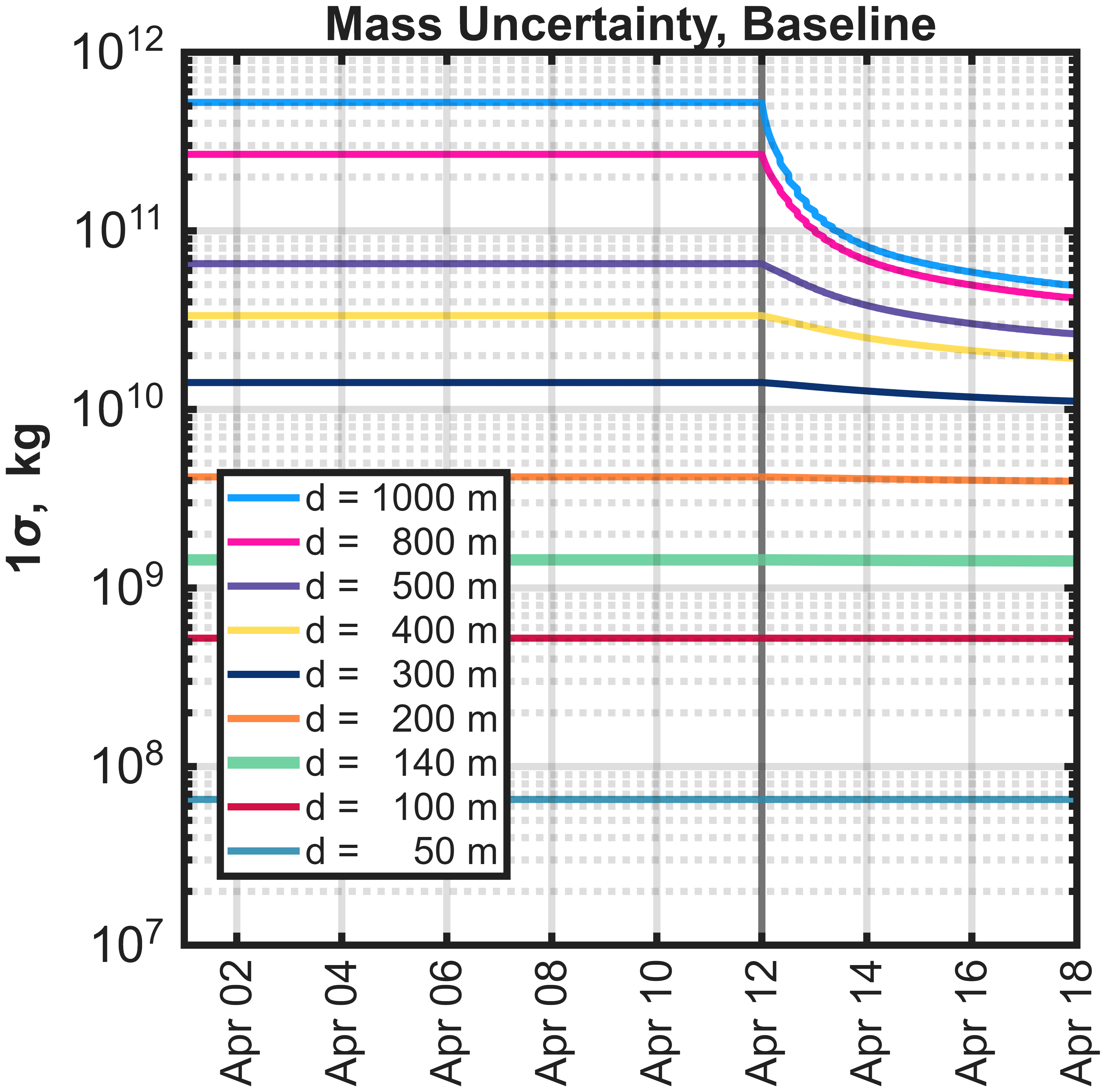} &
  \includegraphics[width=0.475\textwidth]{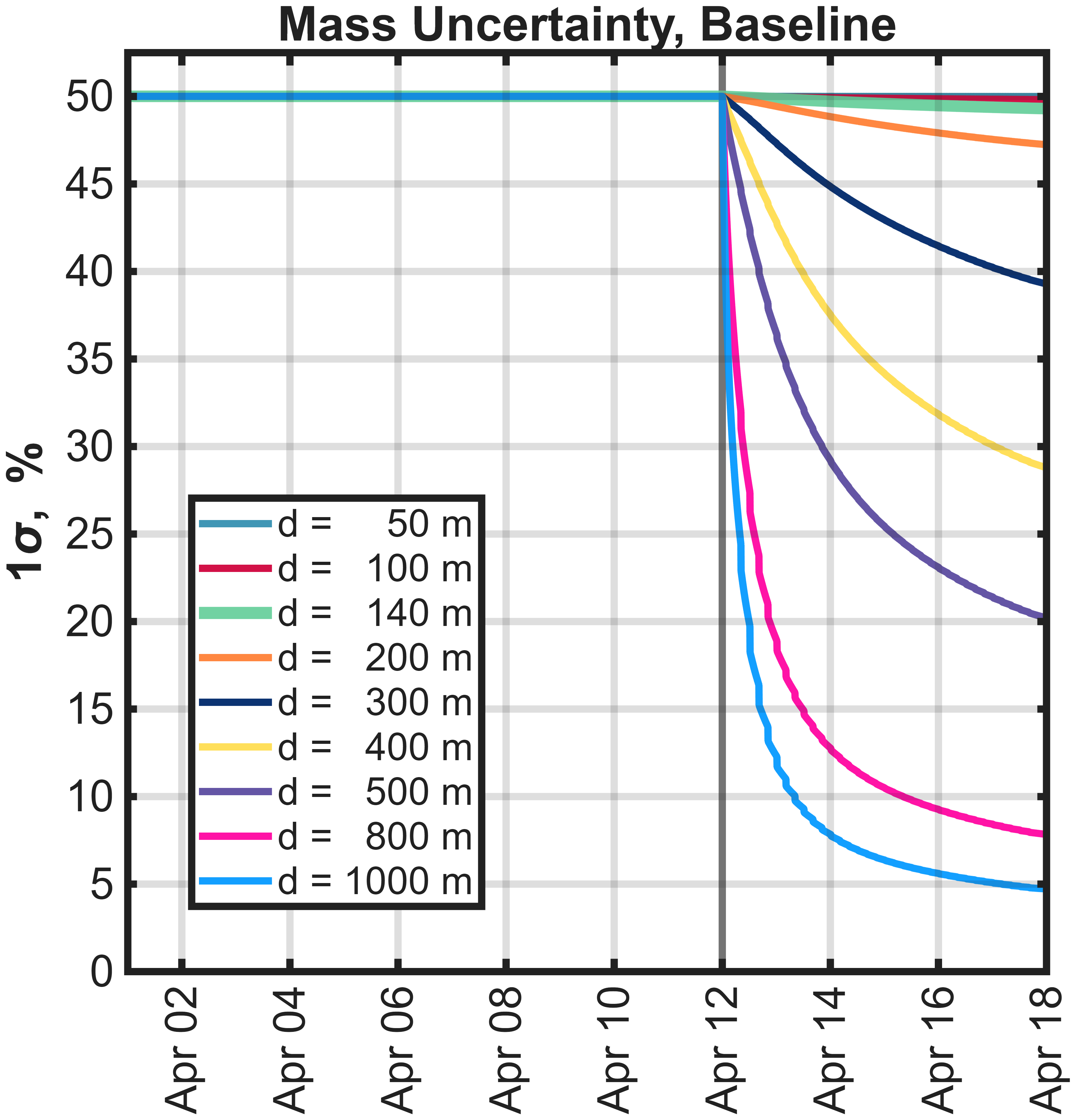} \\
  \\
    (c) & (d) \\
  \includegraphics[width=0.475\textwidth]{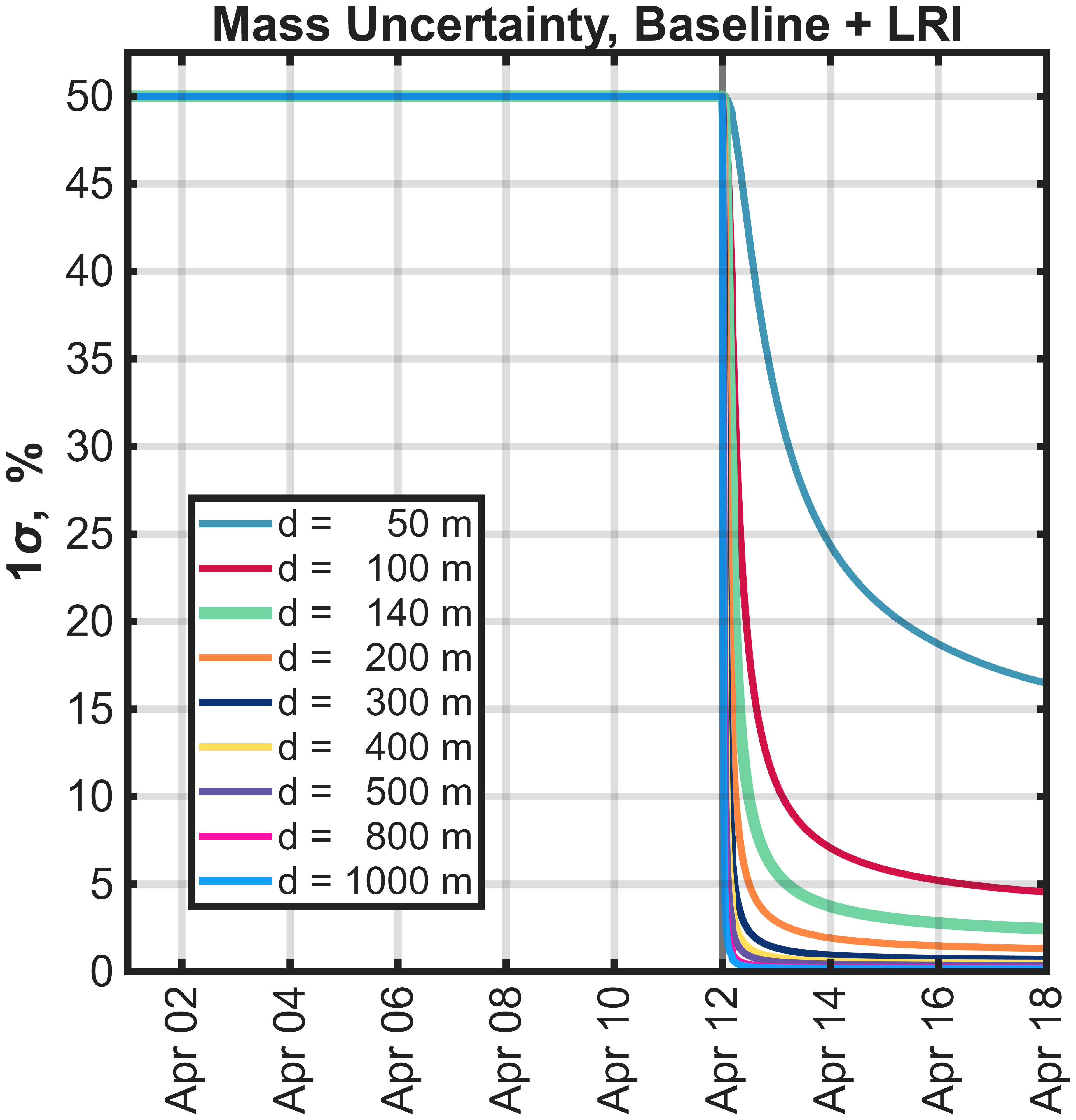} &
  \includegraphics[width=0.475\textwidth]{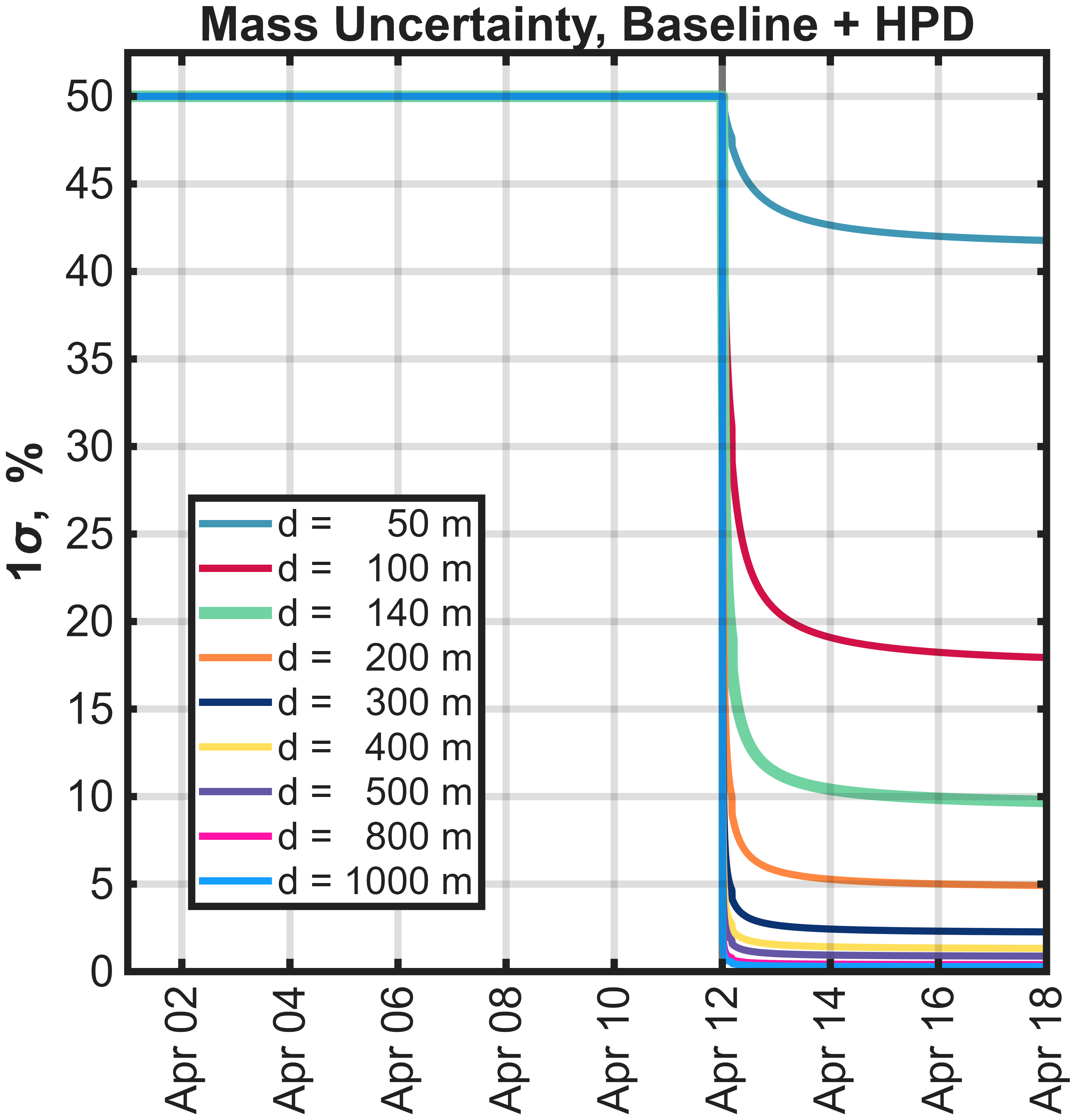} \\
  \\
  \end{tabular}
\caption{
Case 2. Temporal evolution of (a) asteroid mass uncertainty using baseline tracking measurements, (b) asteroid percent mass uncertainty using baseline tracking measurements, (c) asteroid percent mass uncertainty baseline tracking measurements augmented with LRI, and (d) asteroid percent mass uncertainty using baseline tracking measurements augmented with HPD.
\label{fig:case_2_outputs_2}}
\end{figure*}

\begin{figure}
    \centering
    \includegraphics[width=0.6\textwidth]{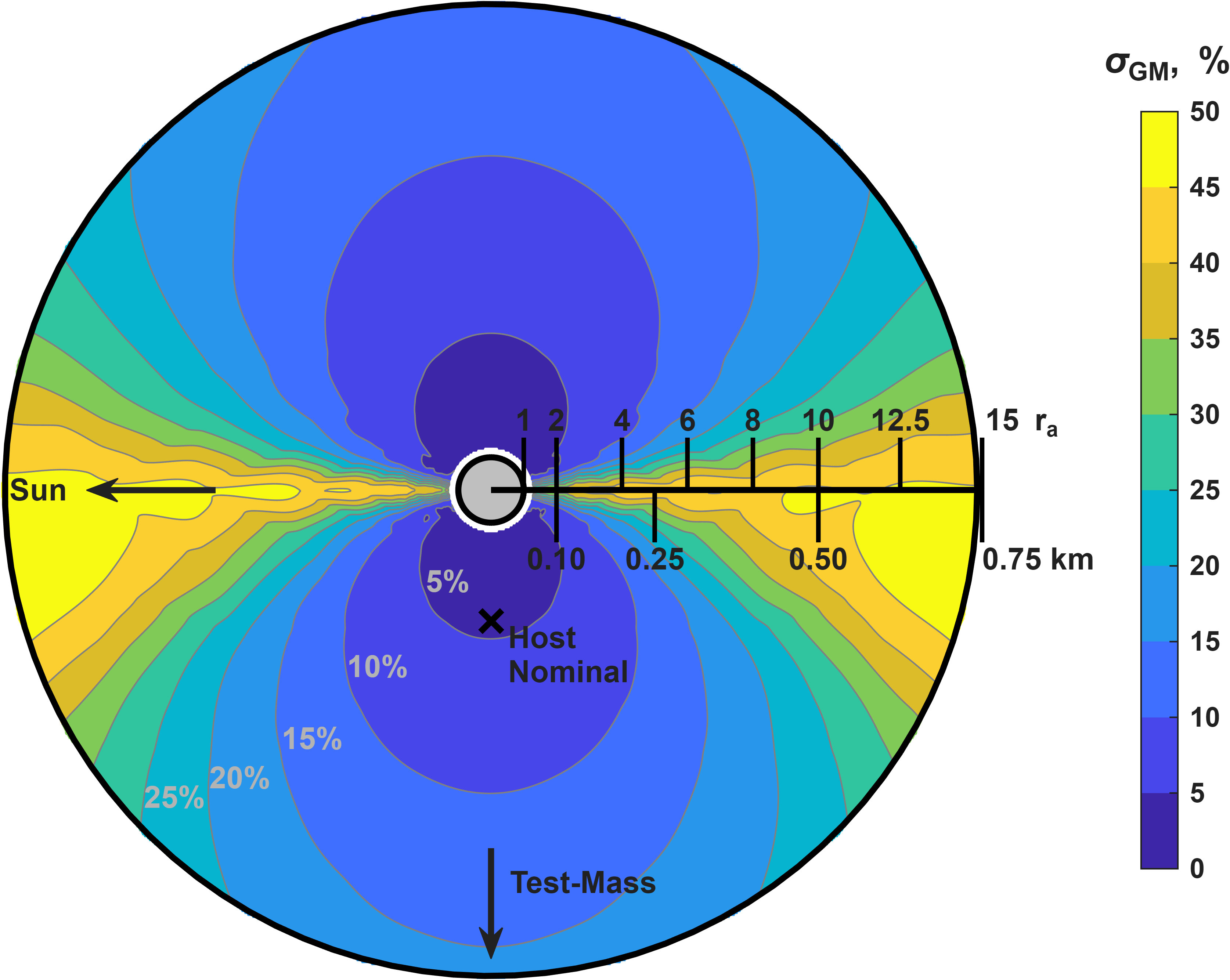}
    \captionof{figure}{Case 2. Final asteroid mass uncertainty as a function of host B-plane flyby location for a 100 m diameter asteroid with LRI augmented measurements}
    \label{fig:2024_PDC25_bpln}
\end{figure}

The results of this medium-speed scenario are given in Figures~\ref{fig:case_2_outputs_1}-\ref{fig:2024_PDC25_bpln}. In this case, the SRP coefficients are less accurately determined over the scenario span compared to the first scenario. The B-plane uncertainty decreases more slowly due to fewer OpNav measurements, although post-flyby it reaches the same 10 m 1$\sigma$ reconstruction knowledge. The final 12 hour host maneuver would occur when the B-plane uncertainty was roughly 600 m 1$\sigma$. This suggests that the low-altitude targeting required for the mass measurement would be particularly challenging operationally. 

Neglecting the B-plane targeting challenge for now, Figures~\ref{fig:case_2_outputs_2}a and b indicate that the baseline measurements can achieve a mass measurement of $1\sigma \leq$ 25\% for asteroids larger than 500 m. When augmented with LRI or HPD, Figures~\ref{fig:case_2_outputs_2}c and d show that this improves to 50 m and 100 m, respectively. 

Figure~\ref{fig:2024_PDC25_bpln} shows the mass uncertainty for the LRI augmented case with a 100 m asteroid as a function of B-plane flyby location. The best performance occurs when the host, test-mass and asteroid all lie along a line, as is our nominal target case. In this case, that line is vertical based on the test-mass's B-plane target. The figure shows that a successful flyby is possible with B-plane errors on the order of roughly 100 m.

\section{Case 3: \yr}\label{sec:case3}
\subsection{Scenario Definition}
The final case we consider is a hypothetical flyby of the real Near-Earth Asteroid \yr. This represents a fast flyby case. \yrs has the distinction of having reached the highest probability of Earth impact, with a value of 3.1\%\footnote[3]{https://cneos.jpl.nasa.gov/sentry/} in mid-February 2025, before dropping to zero with additional observations. At the time of writing it still has a probability of hitting the Moon of 3.8\% in 2032\footnote[4]{https://science.nasa.gov/blogs/planetary-defense/2025/04/02/nasa-update-on-the-size-estimate-and-lunar-impact-probability-of-asteroid-2024-yr4/} and may thus still be an attractive mission target. Prior to the risk of Earth impact becoming negligible \cite{IAWN2025}, we developed rapid reconnaissance flyby options, including the trajectory we consider here. This trajectory launches in May of 2028 and flies past the asteroid 3 months later in August 2028. It is compatible with a ``medium class'' launch vehicle, with no deterministic maneuvers. The flyby speed is 22.1 km/s and the approach solar phase angle spans from 2$^\circ$ to 5$^\circ$ over the 12 days before close approach. While this approach lighting is ideal for OpNav, it means that departure OpNav imaging is not feasible, since the imager would need to point back almost exactly at the Sun. 

\begin{figure*}[tbh!]
  \begin{tabular}{cc}
  \multicolumn{2}{c}{(a)} \\
  \multicolumn{2}{c}{\includegraphics[width=0.5\textwidth]{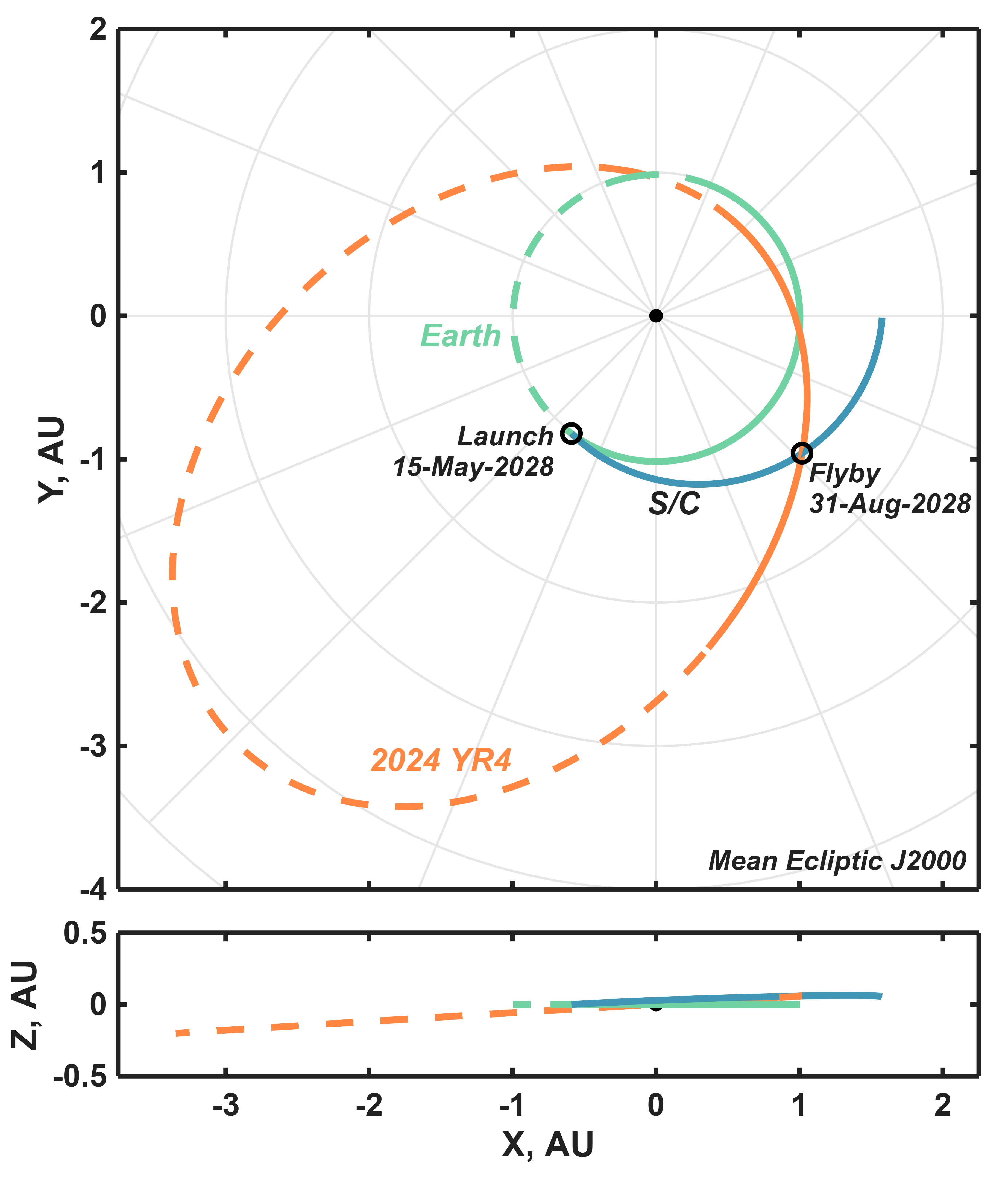}} \\
  \\
    (b) & (c) \\
  \includegraphics[width=0.425\textwidth]{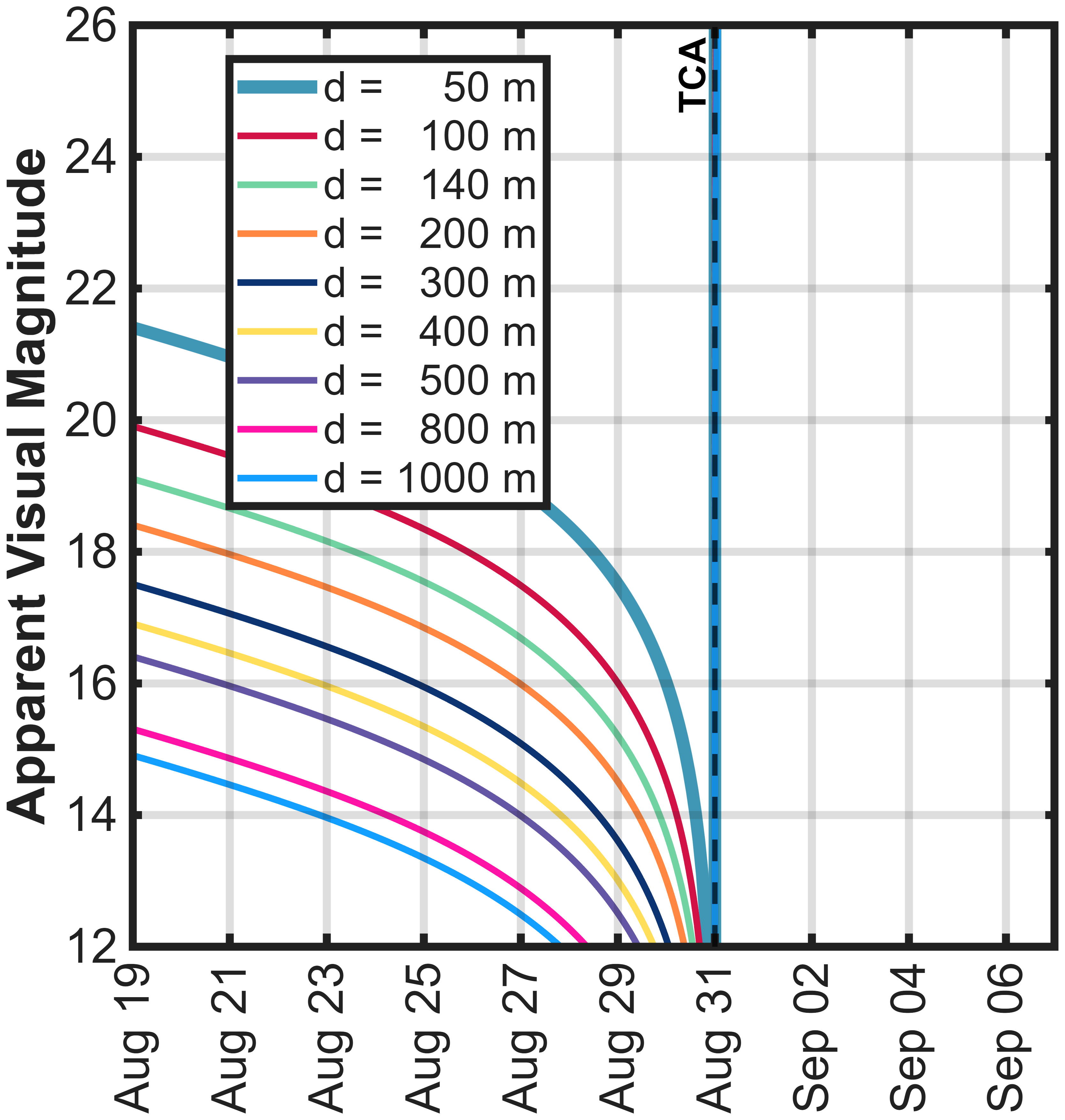} &
  \includegraphics[width=0.475\textwidth]{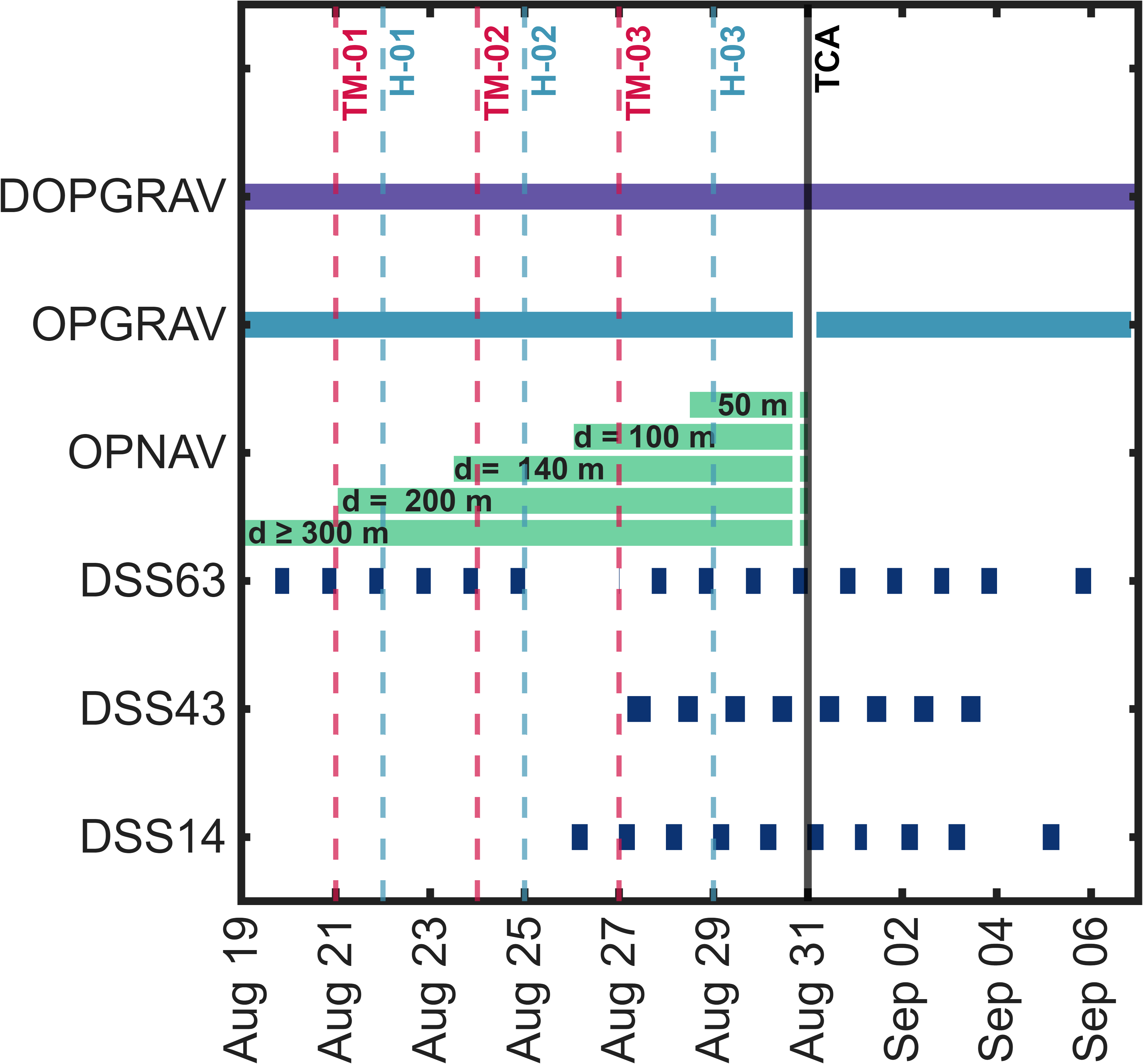}
  \end{tabular}
\caption{
Case 3. (a) Flyby trajectory to \yr. (b) Temporal evolution of the apparent magnitude of \yrs from the spacecraft for varying asteroid diameters. The vertical line indicates the time of closest approach (TCA). (c) Measurement schedule for the encounter with \yr. The vertical dashed lines indicate the times of host (H) and test-mass (TM) maneuvers.
\label{fig:case_3_inputs}}
\end{figure*}

The interplanetary trajectory is shown in Figure~\ref{fig:case_3_inputs}a. The reference diameter of \yrs (60$\pm$7 m) is derived from JWST measurements \cite{rivkin2025}. The apparent magnitude and the nominal measurement schedule are shown in Figures~\ref{fig:case_3_inputs}b \& \ref{fig:case_3_inputs}c. In this case, the OpNav start epoch varies substantially by asteroid diameter.

Both the host and test-mass execute 3 maneuvers to target the encounter geometry. The final test-mass maneuver is $\sim$0.4 m/s and is conducted 4 days prior to close approach. The host's final maneuver is 0.6 m/s and is conducted 12 hours prior to close approach. 

To initialize the asteroid's uncertainty at the start of the scenario, we started with the predicted covariance matrix provided by the Center for Near Earth Object Studies (\yr, Solution \#76) and scaled it to have a maximum $1\sigma$ uncertainty of 100 km in the inertial position coordinates and 0.1 m/s in inertial velocity coordinates. 

\subsection{Results}
\begin{figure*}[tbh!]
  \begin{tabular}{cc}
  (a) & (b) \\
  \includegraphics[width=0.475\textwidth]{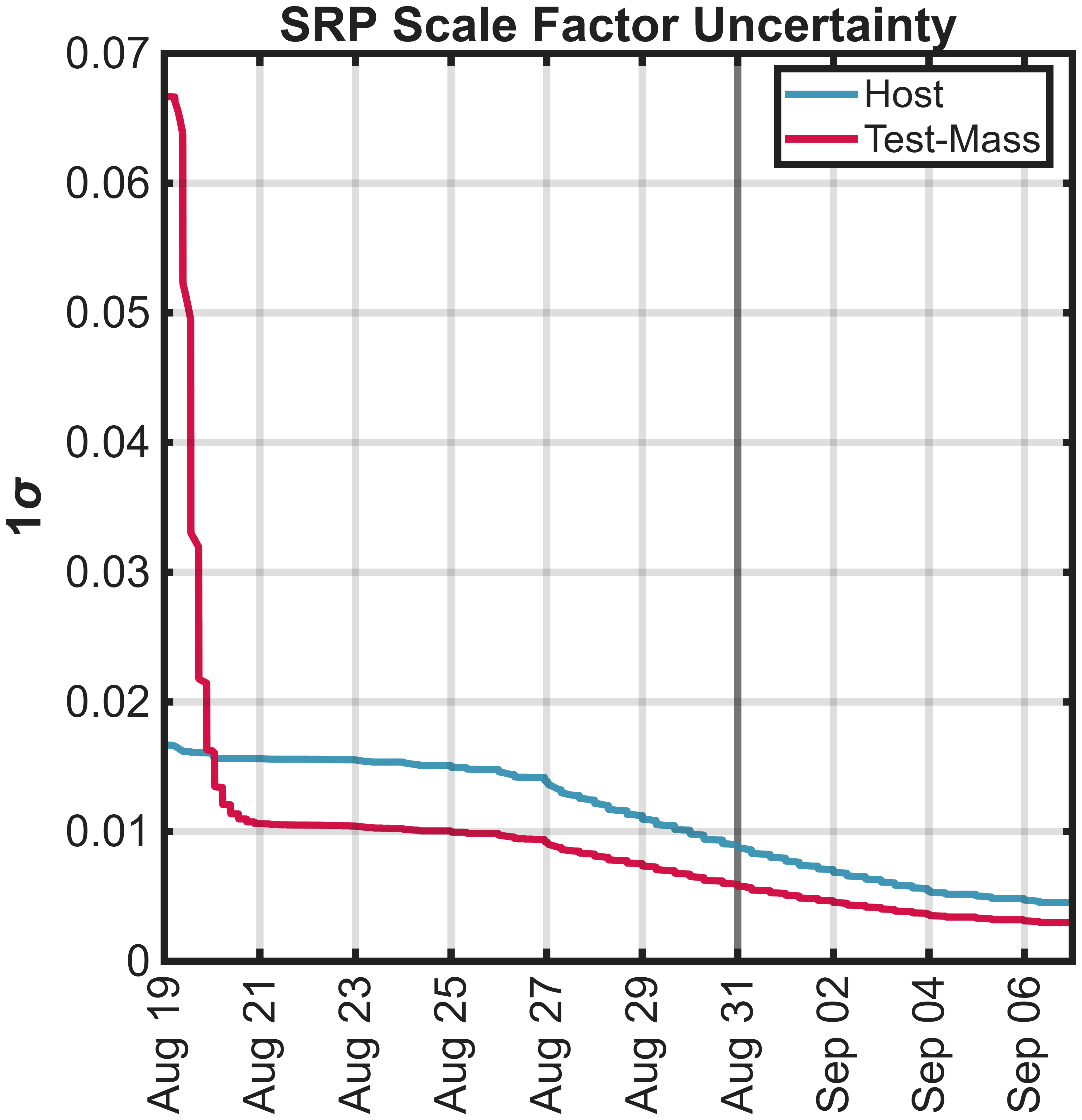} &
  \includegraphics[width=0.485\textwidth]{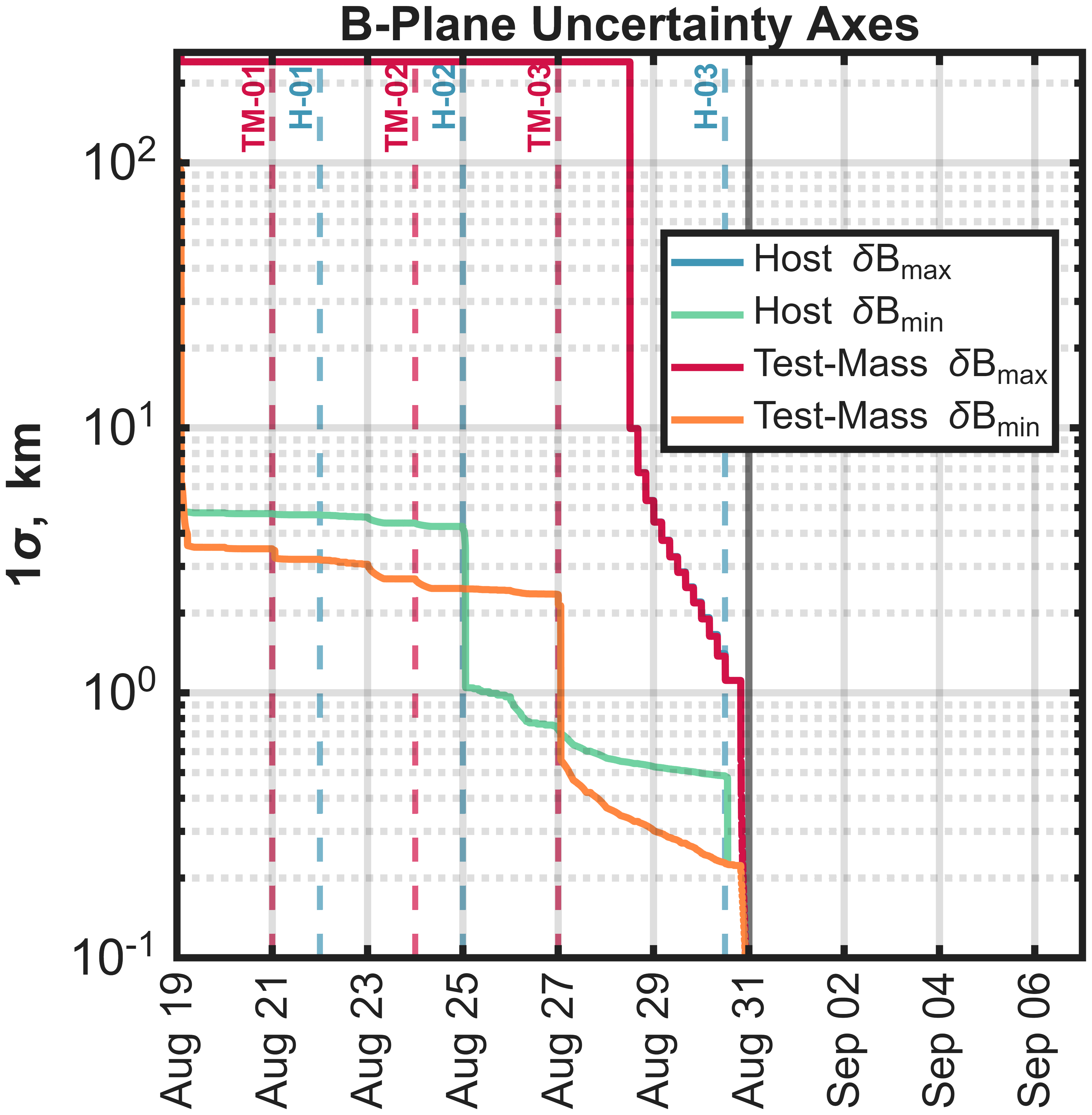} \\
  \\
    (c) \\
  \includegraphics[width=0.485\textwidth]{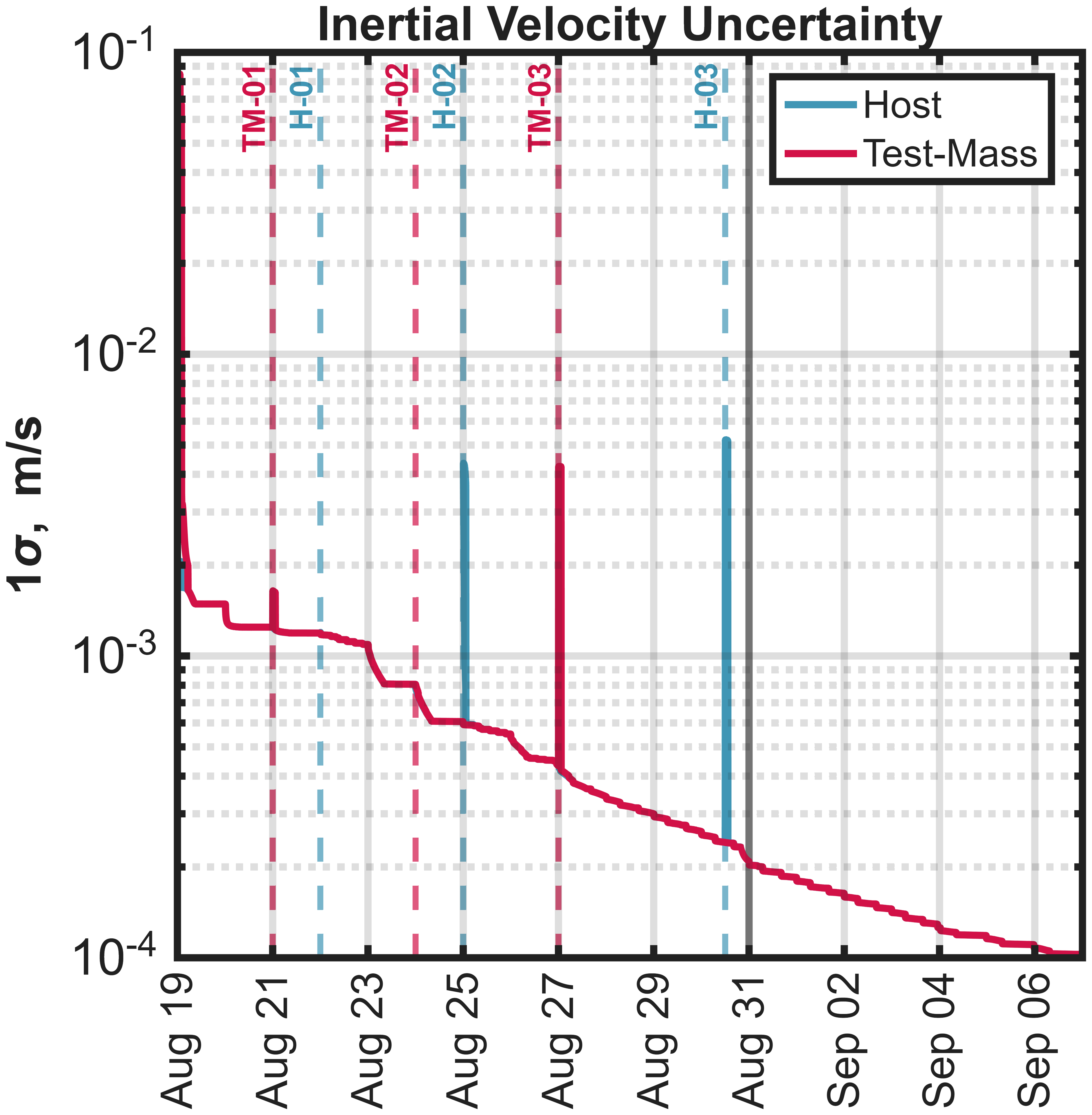} & \\
  \\
  \end{tabular}
\caption{
Case 3. Temporal evolution of spacecraft (a) SRP coefficient uncertainty, (b) B-plane uncertainty ellipse sizes for an encounter with an 50 m diameter asteroid, and (c) inertial velocity uncertainty.
\label{fig:case_3_outputs_1}}
\end{figure*}

\begin{figure*}[tbh!]
  \begin{tabular}{cc}
  (a) & (b) \\
  \includegraphics[width=0.500\textwidth]{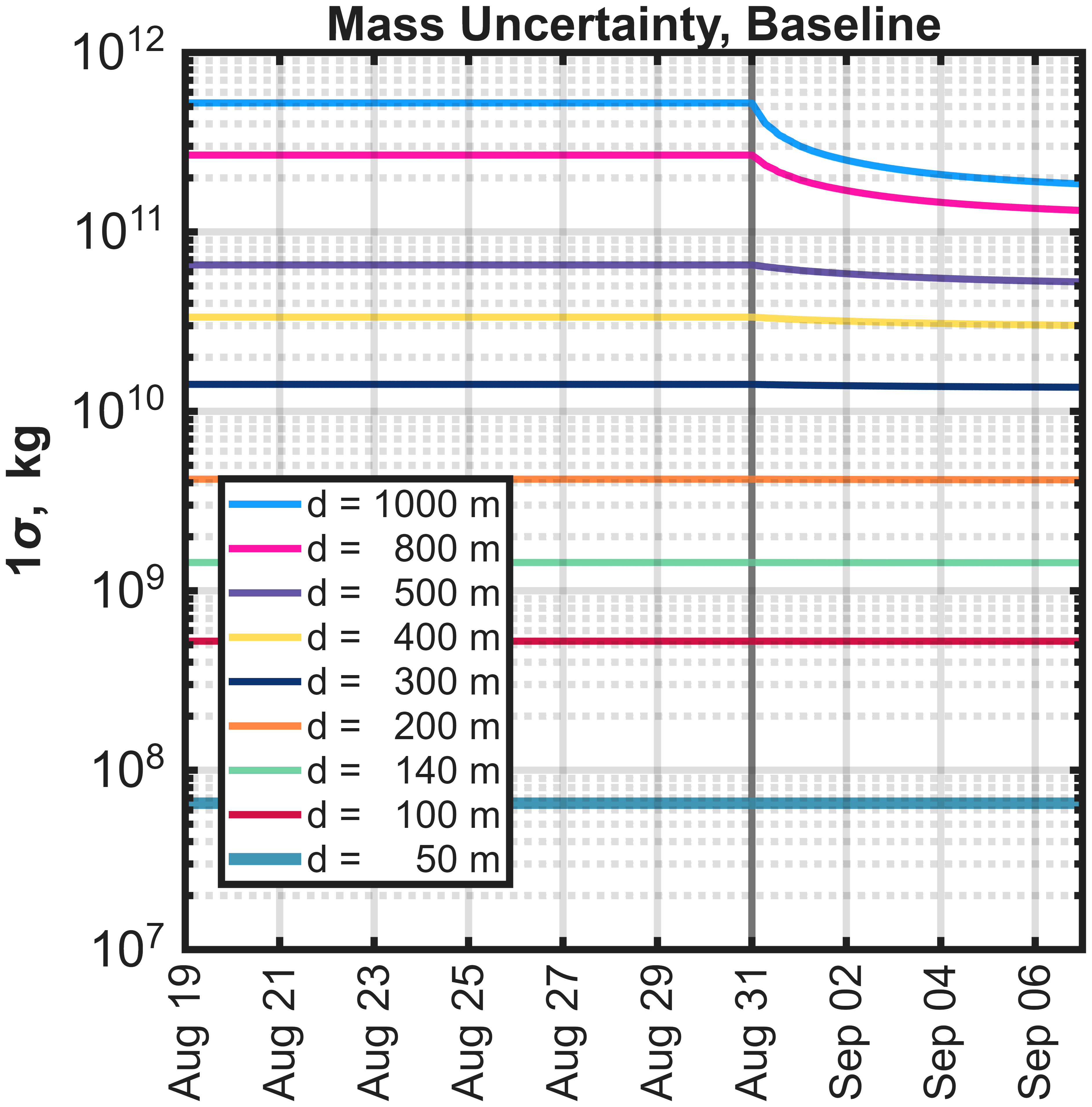} &
  \includegraphics[width=0.475\textwidth]{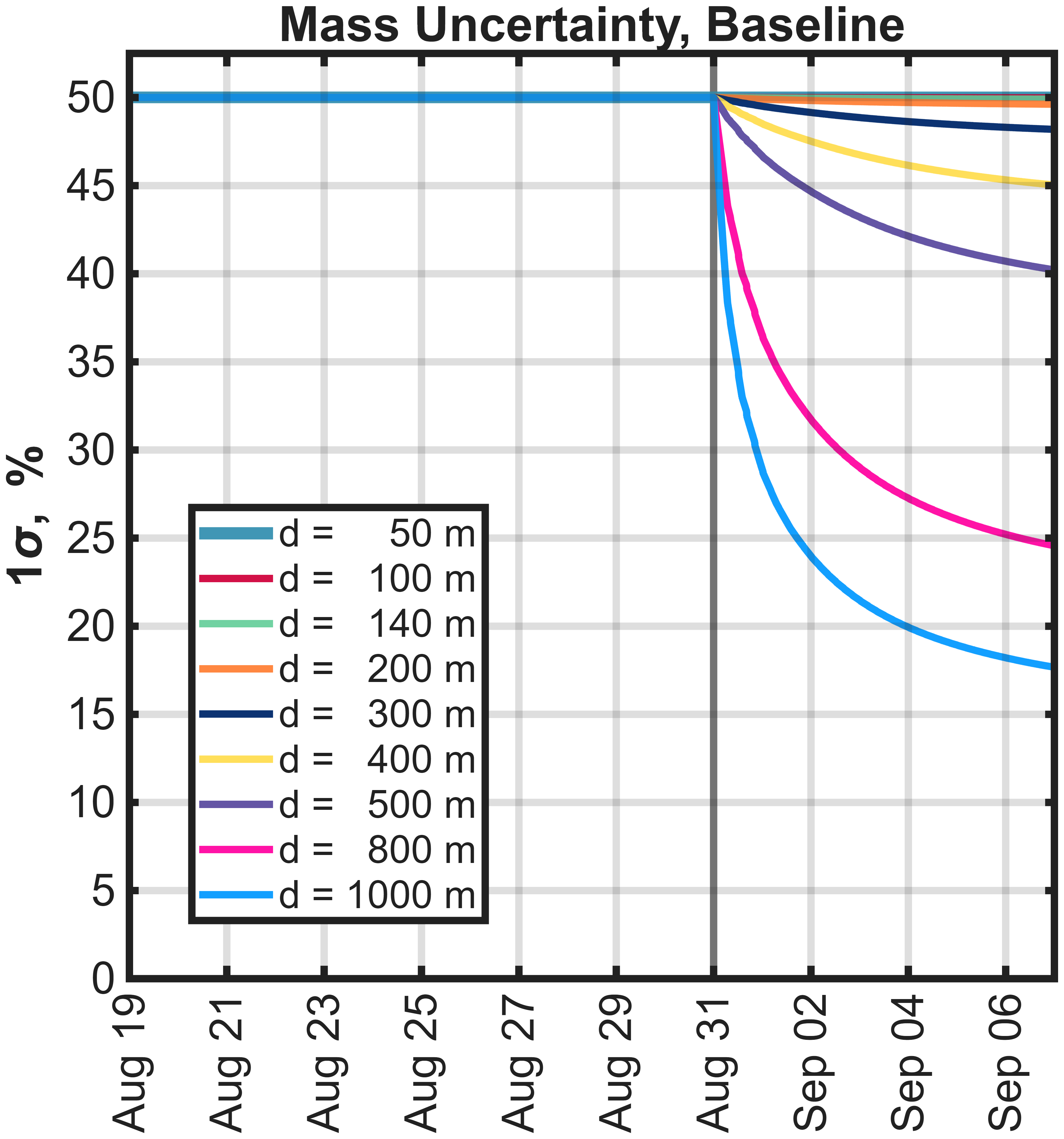} \\
  \\
    (c) & (d) \\
  \includegraphics[width=0.475\textwidth]{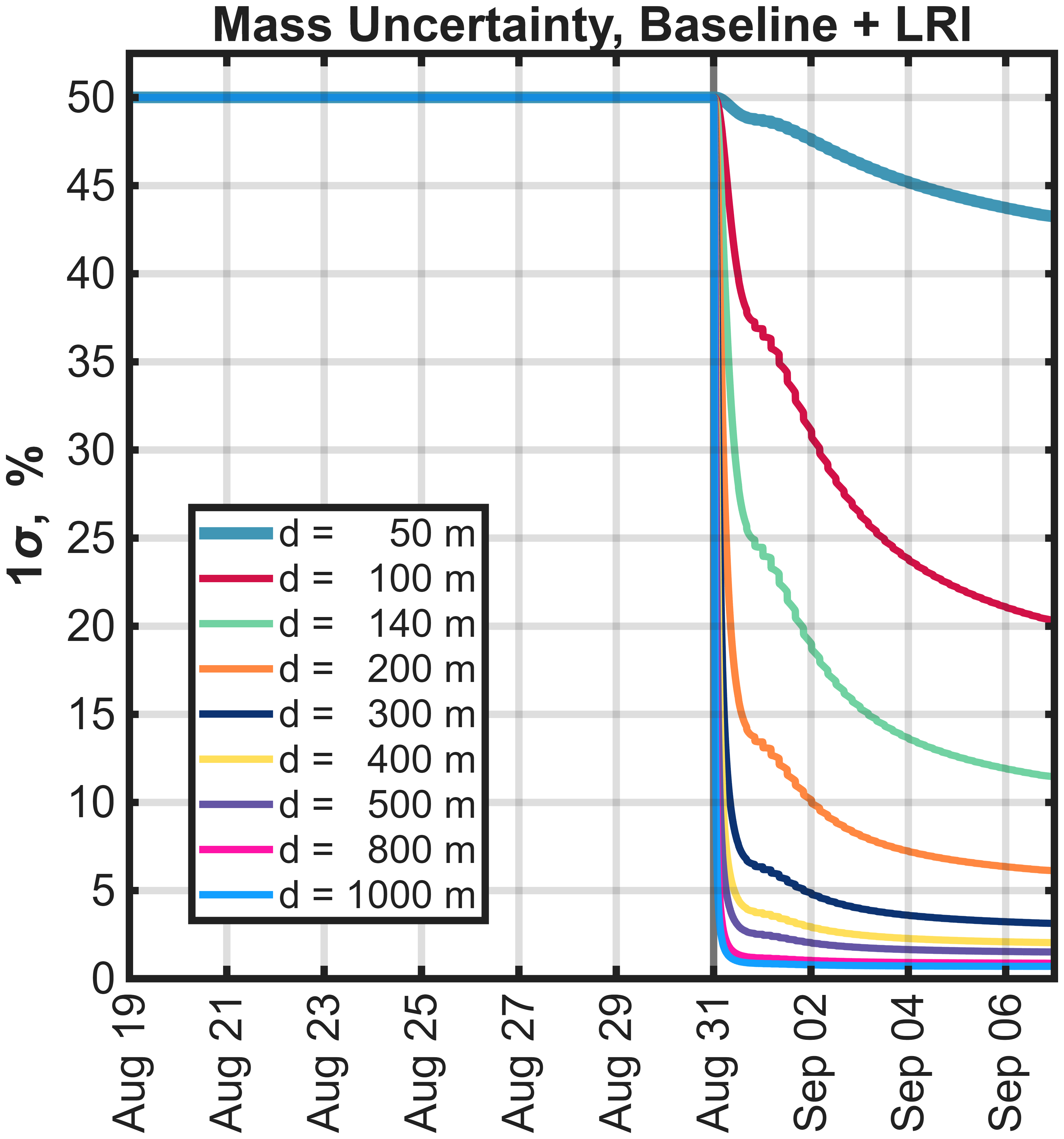} &
  \includegraphics[width=0.475\textwidth]{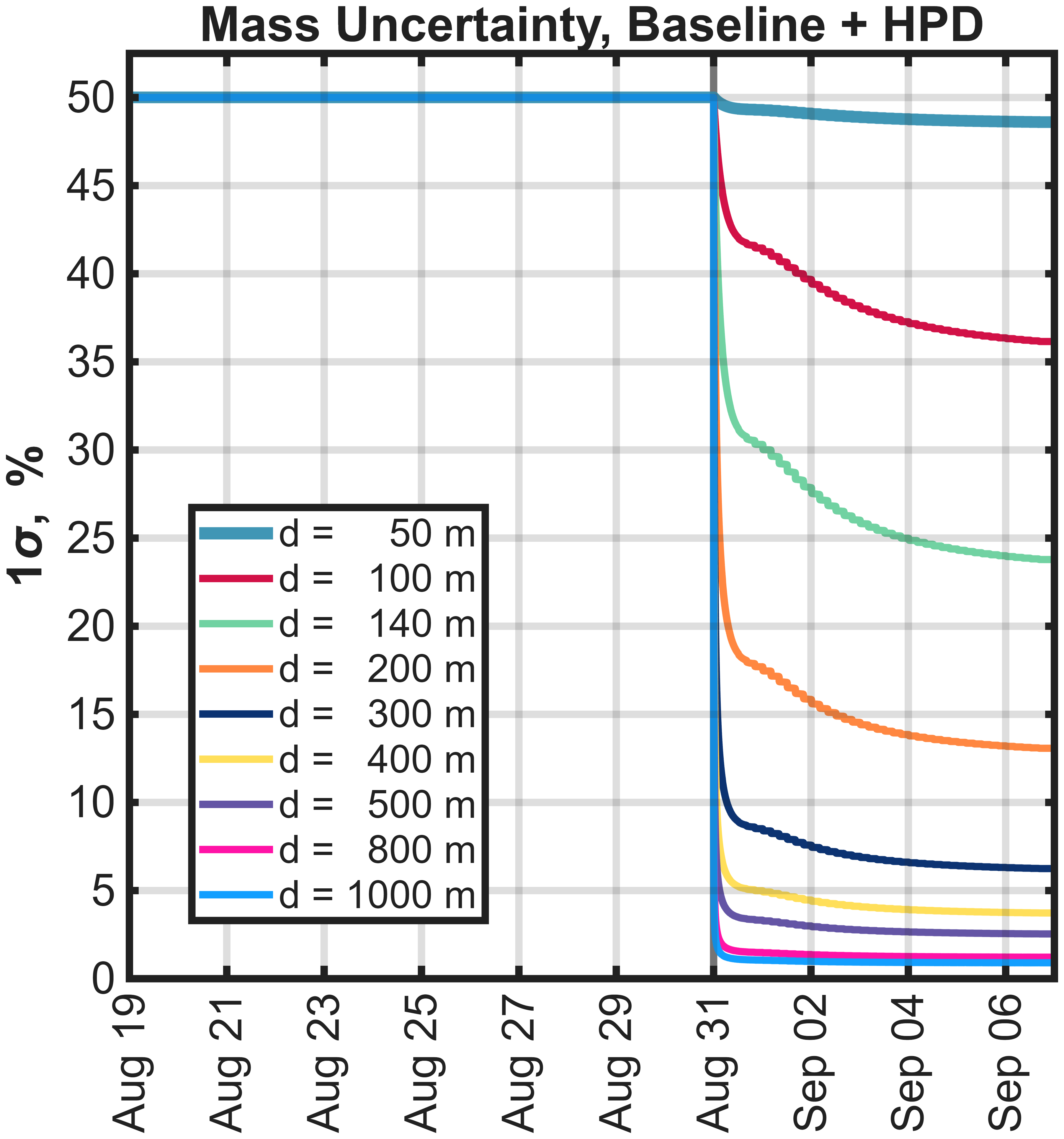} \\
  \\
  \end{tabular}
\caption{
Case 3. Temporal evolution of (a) asteroid mass uncertainty using baseline tracking measurements, (b) asteroid percent mass uncertainty using baseline tracking measurements, (c) asteroid percent mass uncertainty baseline tracking measurements augmented with LRI, and (d) asteroid percent mass uncertainty using baseline tracking measurements augmented with HPD.
\label{fig:case_3_outputs_2}}
\end{figure*}

\begin{figure}
    \centering
    \includegraphics[width=0.6\textwidth]{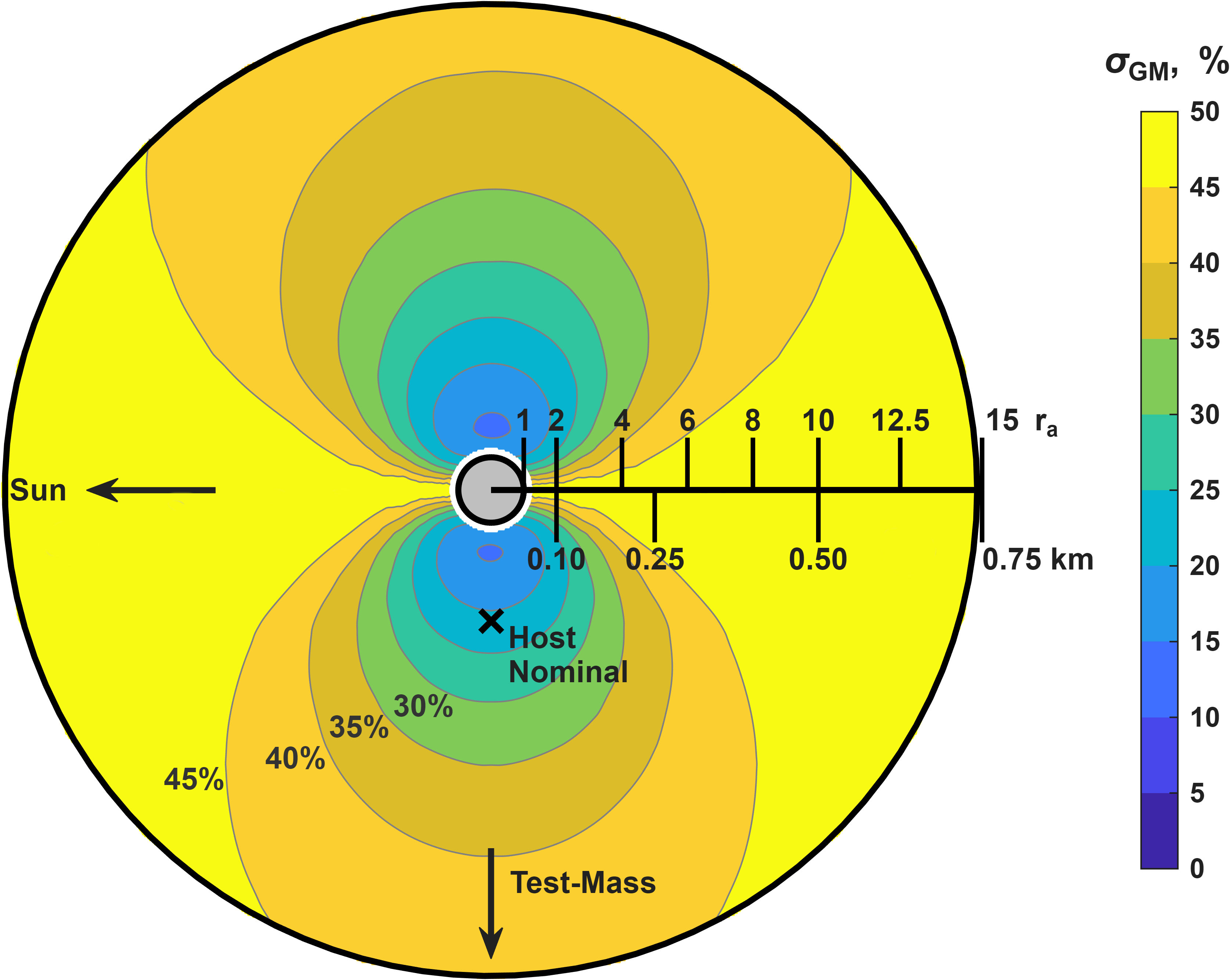}
    \captionof{figure}{Case 3. Final asteroid mass uncertainty as a function of host B-plane flyby location for a 100 m diameter asteroid with LRI augmented measurements}
    \label{fig:2024_YR4_bpln}
\end{figure}

The results for the fast flyby scenario are shown in Figures~\ref{fig:case_3_outputs_1}-\ref{fig:2024_YR4_bpln}. The SRP uncertainty is readily observable, reaching roughly 0.5\% by the end of the scenario. The B-plane uncertainty, however, indicates that the faster approach speed amplifies the operational targeting challenge. The semimajor axis for both the host and test-mass stays at the initial 140 km until OpNav measurements begin. For the 50 m asteroid case shown, this does not begin until 2.3 days prior to the close approach. At the time of the final host maneuver, the B-plane uncertainty is  $\sim$1.5 km x 0.5 km (1$\sigma$), or approximately $\sim$60 x 20 body radii. This indicates that the scenario would not be operationally feasible. There is insufficient B-plane knowledge to ensure a safe flyby of the asteroid, or even design the final targeting maneuver. As in the prior cases, there is sufficient data to accurately reconstruct the true B-plane location after the flyby down to roughly 10 m $1\sigma$.

The mass measurement performance is stressed for the baseline measurements, with a mass measurement only possible for only the 1 km asteroid. When LRI and HPD are included, the performance is substantially improved, showing successful flybys for asteroids as small as 100 m and 140 m respectively.

Figure~\ref{fig:2024_YR4_bpln} shows the sensitivity of the mass measurement to B-plane flyby location for the LRI equipped host flying past a 100 m asteroid. The dependence on flyby altitude is apparent, with the nominal uncertainty ($\sim$20\%) doubling ($\sim$40\%) at a flyby altitude of 11 body radii (550 m), compared to the nominal flyby altitude of 3 body radii (150 m).   

\begin{figure*}[tbh!]
  \begin{tabular}{cc}
  (a) & (b) \\
  \includegraphics[width=0.475\textwidth]{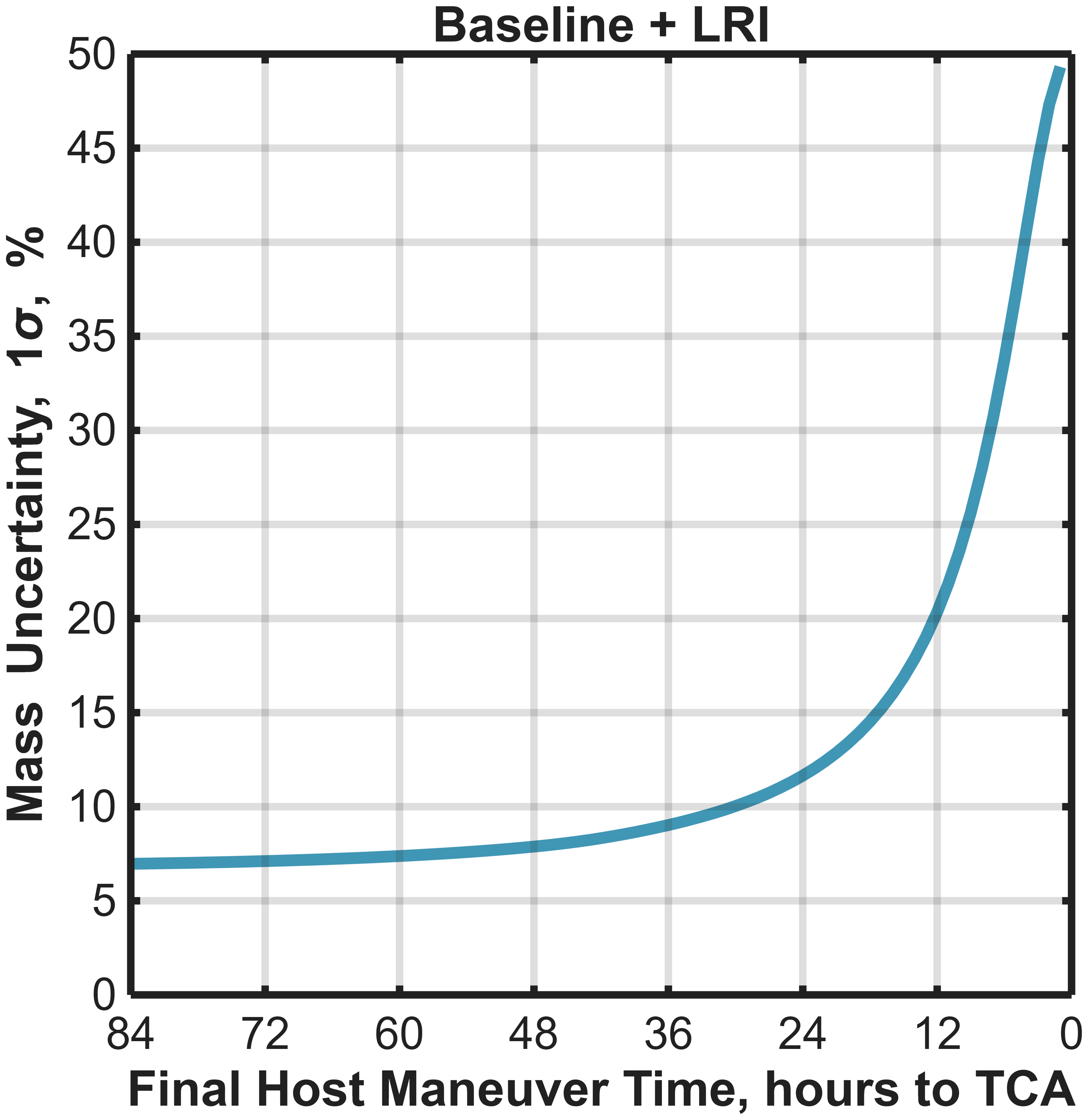} &
  \includegraphics[width=0.485\textwidth]{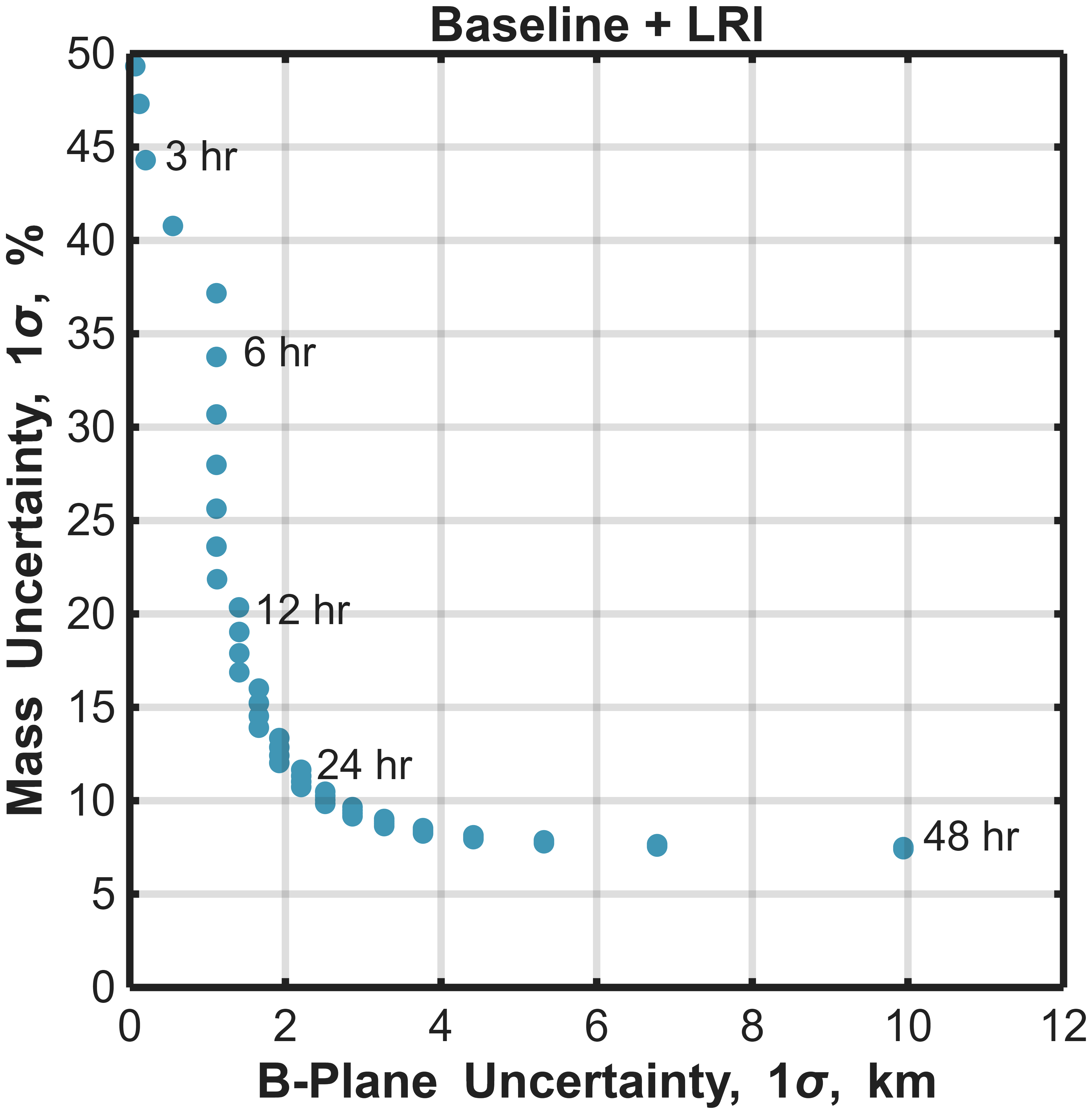} \\ 
  \\
  \end{tabular}
\caption{
Case 3. (a) Mass uncertainty as a function of final host maneuver time, for a 100 m diameter asteroid with LRI augmented measurements. (b) Mass uncertainty as a function of B-plane uncertainty at the time of final maneuver, for a 100 m diameter asteroid with LRI augmented measurements.
\label{fig:case_3_outputs_3}}
\end{figure*}

One approach for improving the B-plane targeting uncertainty is to maneuver closer to the time of the encounter, when the B-plane uncertainty is further reduced. In an effort to understand the consequences of this on mass performance and identify an operationally feasible timeline, we parametrically evaluated a range of final maneuver times. Figure~\ref{fig:case_3_outputs_3}a shows the final mass measurement 1$\sigma$ accuracy for these cases, for a 100 m asteroid with LRI. The nominal case, with a maneuver at 12 hours, has a B-plane uncertainty of 1.4 km and a mass uncertainty of roughly 20\%. Moving the maneuver time to 6 hours later decreases the B-plane uncertainty to 1.1 km, but has a substantial negative effect on the asteroid's mass uncertainty, increasing it to 34\%. If the maneuver could be conducted earlier at 72 hours before close approach, when the B-plane uncertainty is 10 km (ignoring the aforementioned operational challenges), the mass uncertainty could hypothetically be improved to roughly 7\%. Note that this analysis neglects the time required to downlink and process OpNav measurements on the ground and subsequently upload a targeting maneuver (the data cutoff). It shows the performance for an instantaneous orbit determination update and maneuver design on-board the spacecraft.

Figure~\ref{fig:case_3_outputs_1}b shows the trade-space between the final potentially achievable mass uncertainty (if the host is successfully delivered to its target B-plane point) and the B-plane uncertainty at the time of the final maneuver's execution. If we wait to conduct the maneuver when the B-plane uncertainty is less than 100 m (roughly 3 hours before close approach), the mass measurement is effectively negated.

This parametric study demonstrates a fundamental challenge for high flyby speeds. To achieve a small close approach distance, we need a small B-plane uncertainty. B-plane uncertainty reduces with distance to the asteroid, so we want to delay the final maneuver as long as possible. However, the uncertainty of a late maneuver can persist through close approach and consequently corrupt or overwhelm the mass measurement. At lower flyby speeds, with sufficient detection time, we can find a final maneuver time that satisfies both constraints. For higher flyby speeds, we need to explore means of detecting the asteroid earlier to increase the number of available OpNav measurements and reduce B-plane uncertainties sooner.

\section{Summary of Results}\label{sec:summary}
The results across the three cases are summarized below in Figure~\ref{fig:rollup_outputs_1} and Table~\ref{tab:results}. Figure~\ref{fig:rollup_outputs_1}a shows the baseline results, which shows that it is only suitable for relatively large threats and/or slower flyby speeds. The black points indicate the scenario's reference diameter from Table~\ref{tab:scenario_list}. Figure~\ref{fig:rollup_outputs_1}b and c show the baseline augmented with the LRI and HPD respectively. 

\begin{figure*}[h!]
  \begin{tabular}{cc}
  (a) & (b) \\
  \includegraphics[width=0.475\textwidth]{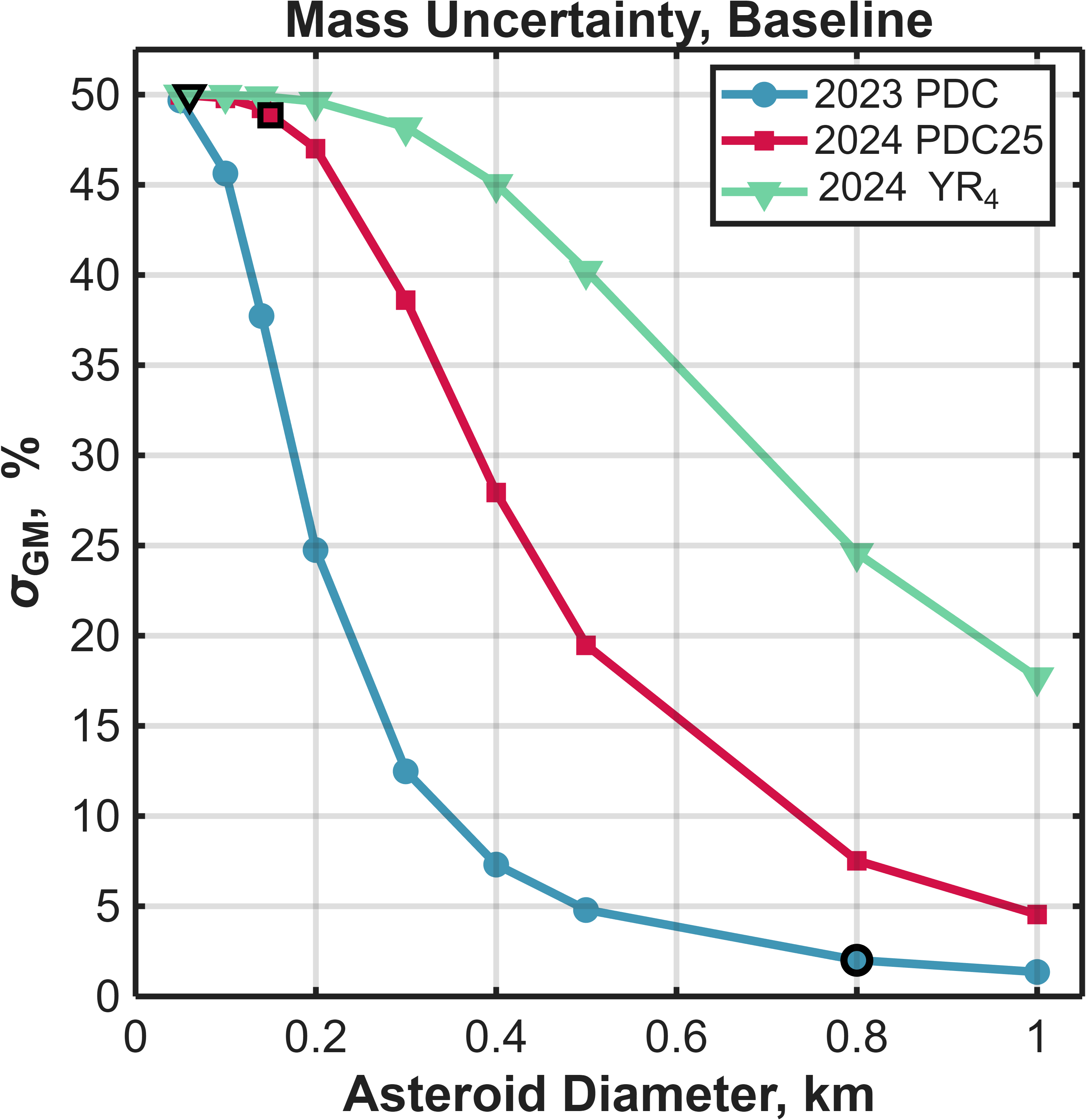} &
  \includegraphics[width=0.475\textwidth]{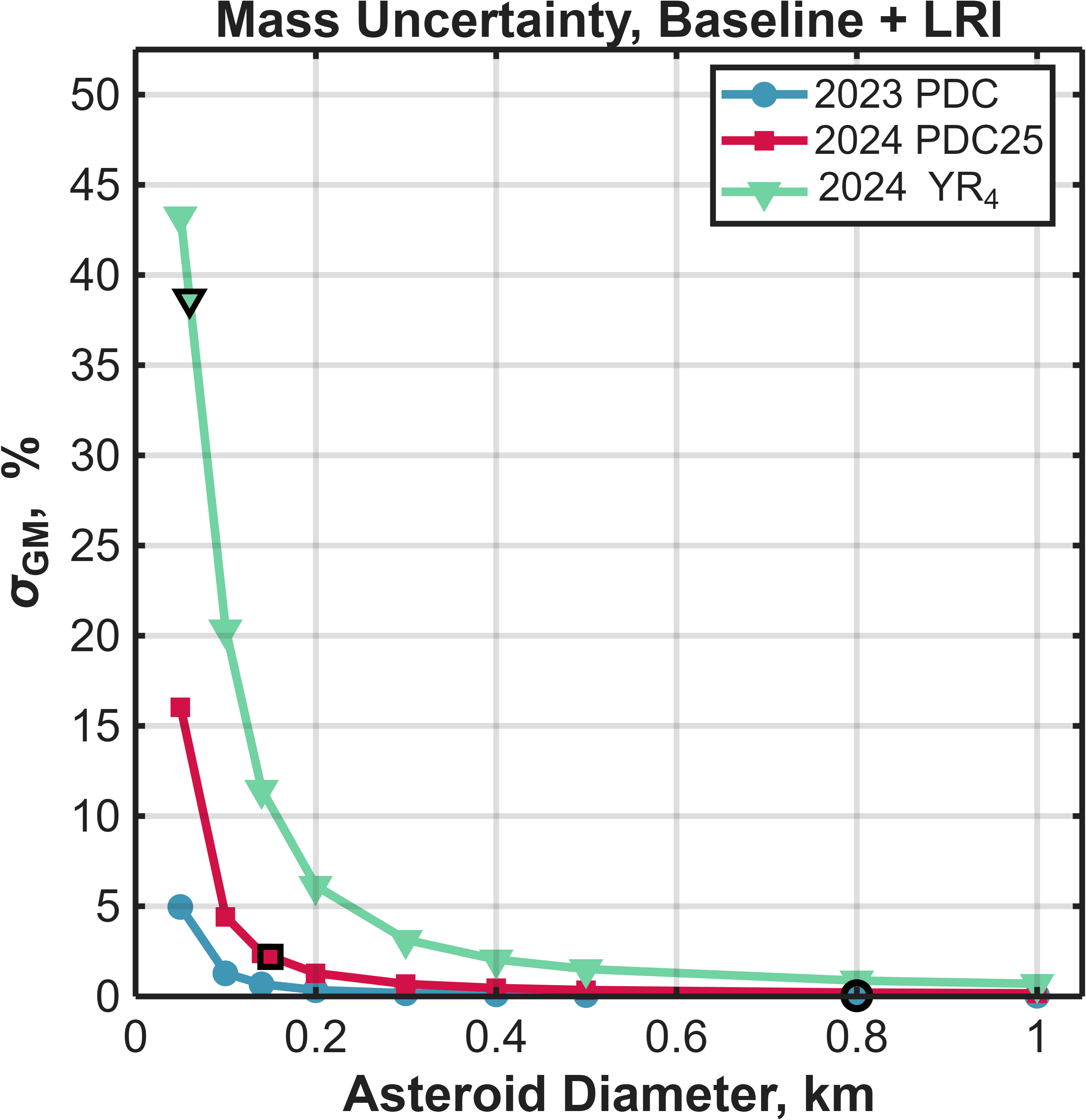} \\
  \\
    (c) & (d) \\
  \includegraphics[width=0.475\textwidth]{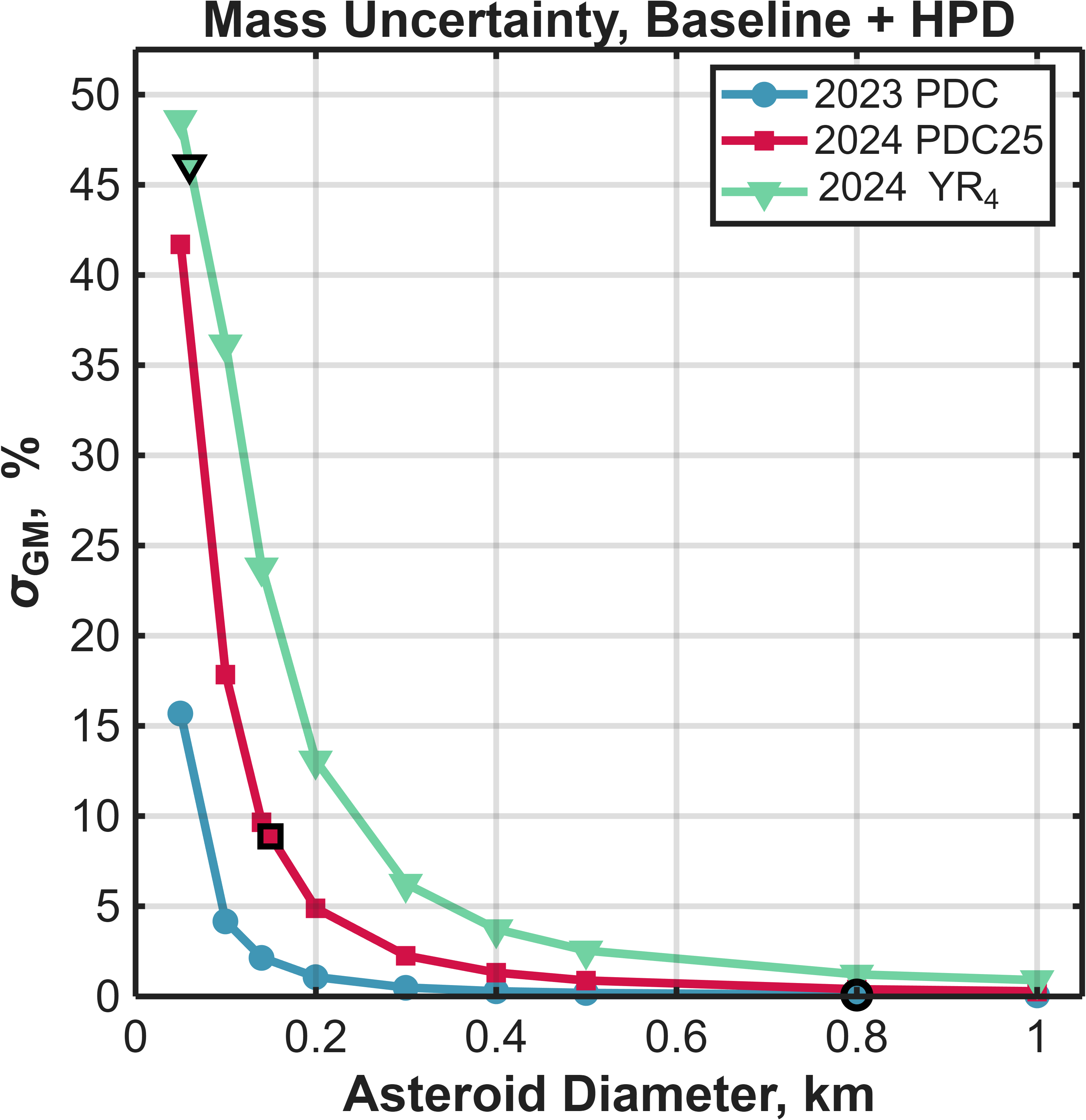} &
  \includegraphics[width=0.475\textwidth]{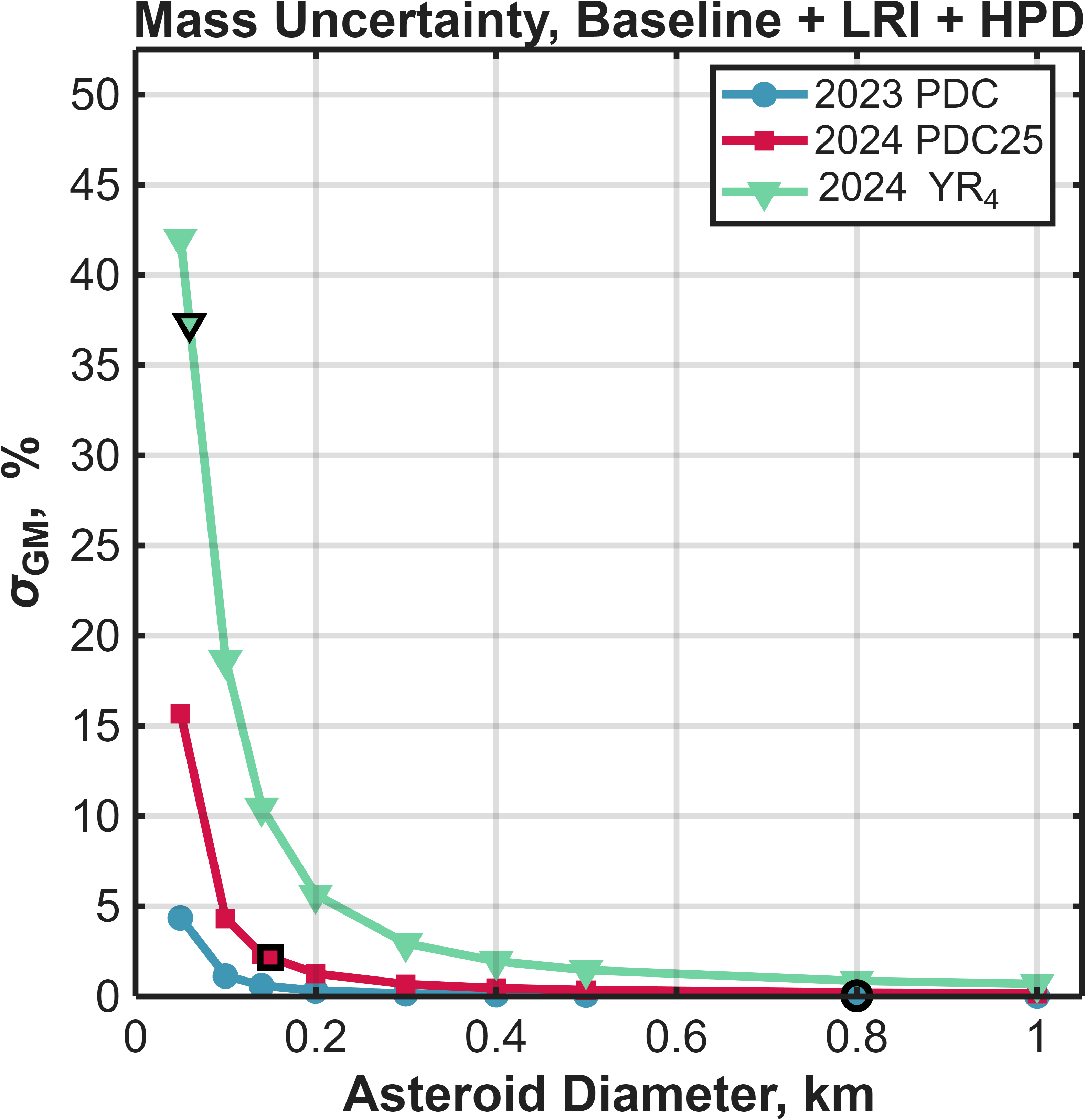} \\
  \\
  \end{tabular}
\caption{
Final mass uncertainty for all 3 scenarios, using: (a) baseline tracking measurements (b) baseline tracking measurements augmented with LRI, (c) baseline tracking measurements augmented with HPD, (d) baseline tracking measurements augmented with both LRI and HPD.
\label{fig:rollup_outputs_1}}
\end{figure*}

When using either LRI or HPD, asteroids as small as 50 m are measurable for the 2023 PDC and 2024 PDC25 cases, but not for the \yrs case with the fastest flyby speed.  Figure~\ref{fig:rollup_outputs_1}d shows the best possible scenario of equipping the host spacecraft with \textit{both} LRI and HPD. This case shows only marginal improvement over the LRI case. 

\begin{table}
    \centering
    \caption{\small{Summary of Results}}
    \label{tab:results}
    \begin{tabular*}{0.95\columnwidth}{lrrr}
        \toprule
        & \multicolumn{3}{c}{Smallest Asteroid Diameter Measurable with $\leq$25\% Uncertainty} \\
        \midrule
        Measurements    &   Case 1: 2023 PDC & Case 2: 2024 PDC25 & Case 3: \yr \\
         \midrule
        Baseline        &    200 m            & 450 m               & 800 m\\
        Baseline + LRI  &  $<$50 m            & $<$50 m             & 100 m\\
        Baseline + HPD  &  $<$50 m            & 80 m                & 160 m\\
        Baseline + LRI + HPD  &  $<$50 m      & $<$50 m             & 80 m\\
        \bottomrule
    \end{tabular*}
\end{table}

\section{Conclusions}\label{sec:conclusions}
This study demonstrates that a pair of spacecraft can potentially achieve sensitive measurements of an asteroid's mass from a high speed reconnaissance flyby. 

The baseline radiofrequency (RF) intersatellite measurements do not achieve a meaningful mass measurement for asteroids as small as 140 m in diameter. However, we found that this goal is possible if we either can reduce the intersatellite range accuracy to roughly 1 cm (1$\sigma$) using a laser ranging instrument (LRI) or reduce the intersatellite range-rate accuracy to 0.1 $\mu$m/s using a high precision Doppler (HPD) instrument with an ultrastable oscillator. 

The prevailing challenge we identified is to find an operational B-plane targeting timeline consistent with the expected approach knowledge. For the smallest asteroids, the host spacecraft needs to achieve a very low flyby distance, with errors on the scale of 100 m or less. And for high-speed flybys (\textit{e.g.} $>$10 km/s), this accuracy is not available until the final hours of approach. We find that if we conduct a targeting maneuver within this time-frame (\textit{e.g.} $\sim$6 hours of close approach), the mass measurement is substantially degraded if not negated. This is true even for very small maneuvers of only a few cm/s. 

Going forward, we will study the OpNav knowledge question further, seeking to identify potential solutions. For example, we might consider one of the following:
\begin{itemize}
    \item At worst, it may be possible to constrain the approach speed as a function of asteroid size and detectability. 
    \item We may be able to select or design a more sensitive camera or incorporate longer integration times. This would allow us to detect the asteroid sooner and incorporate earlier OpNav measurements. 
    \item We may be able to detect the asteroid sooner by co-adding images on-board to achieve a higher sensitivity. 
    \item With further on-board processing, we may be able to improve the rate of uncertainty reduction by incorporating nearly continuous OpNav measurements.
    \item We could consider sending an OpNav ``scout'' spacecraft, which could conduct a relatively distant (\textit{e.g.} 10 or 100 km) flyby the asteroid multiple days in advance of the gravity-science spacecraft.
    \item We might consider sending multiple low altitude spacecraft to further increase our sensitivity in spite of late measurements (\textit{i.e.}, \cite{bull2025}).
    \item And finally, the availability of a future deep space radar \cite{taylor2020} capable of providing an improved pre-encounter asteroid uncertainty may help mitigate this challenge.
\end{itemize}  

If this challenge can be addressed, and if a LRI or HPD instrument is incorporated, our simulations show that a very precise mass measurement is possible from flyby, even at objects as small as 50 m. This represents an important reconnaissance capability in a real-world planetary defense threat scenario. 

\section{Acknowledgements}
The authors wish to thank Davide Farnocchia and Paul Chodas of JPL for providing the asteroid \textit{a-priori} covariances and Jim Woodburn of Ansys for software support with ODTK.

\bibliographystyle{IEEEtran}
\bibliography{references}

\end{document}